\documentclass[aps,prb,twocolumn,superscriptaddress,longbibliography]{revtex4-2}

\usepackage{amsmath,amssymb,graphicx,bm}
\usepackage[colorlinks=true,linkcolor=blue,citecolor=blue,urlcolor=blue]{hyperref}

\newcommand{\bk}{\mathbf{k}}

\newcommand{\bR}{\mathbf{R}}

\newcommand{\mm}{\hat{\bm{m}}}
\newcommand{\btau}{\bm{\tau}}
\newcommand{\Jt}{\bm{J}}
\newcommand{\Am}{\bm{A}}
\newcommand{\Ah}{\hat{\bm{A}}}
\newcommand{\Dc}{D}
\newcommand{\eps}{\varepsilon}

\begin{document}

\title{Classical dipolar Heisenberg models on the Archimedean and Laves
lattices}

\author{J.~Batle}
\affiliation{CRISP -- Centre de Recerca Independent de sa Pobla,
             sa Pobla, Balearic Islands, Spain}
\affiliation{Departament de F\'{\i}sica, Universitat de les Illes Balears,
             07122 Palma de Mallorca, Balearic Islands, Spain}

\date{\today}

\begin{abstract}
We study classical Heisenberg spins interacting through the
dipole--dipole interaction on the eleven Archimedean and eight
Laves planar lattices at canonical geometry, which reduce to fifteen
distinct vertex sets.
For each lattice we compute the Fourier-space interaction matrix
$\Am(\bk)$, extract the Luttinger--Tisza ordering wave vector $\bk_{0}$,
determine the classical ground state by unconstrained minimization on the
commensurate magnetic cell, and evaluate the linearized
spin-wave dispersion $\varepsilon_{i}(\bk)$ along the high-symmetry
path of the (magnetic) Brillouin zone together with the
Holstein--Primakoff moment reduction.
The ground states fall into three classes: five are collinear, two are
non-collinear at angles commensurate with the lattice symmetry, and eight
cant at transcendental angles, which we determine to thirty digits.
Across the family, failure of the Luttinger--Tisza strong condition
coincides exactly with the appearance of incommensurate canting, with no
exceptions in either direction. Truncating the interaction range shows that
whether a lattice cants is fixed by the local coordination geometry, whereas
the value of the canting angle is set by the long-range tail. The quantum
corrections are governed by the magnon spectrum through the size of the
magnetic basis and are uncorrelated with frustration, so that the frustration
and fluctuation classifications are independent; the corner-sharing lattices
host the narrowest low-energy branches.
The results constitute a spectral atlas of dipolar Heisenberg systems
on the full family of $1$-uniform planar tilings and their duals, and
provide predictions for inelastic neutron scattering in materials
whose magnetic ions occupy Archimedean or Laves lattices, and for
artificial arrays of dipolar-coupled nanomagnets.
\end{abstract}

\maketitle

\section{Introduction}\label{sec:intro}

The dipole--dipole interaction is unavoidable in magnetic matter. It is
usually subdominant to exchange, but whenever exchange is weak, strongly
screened, or frustrated to near-cancellation, the dipolar term becomes the
residual coupling that selects the ordered state
\cite{DeBell2000,Wessler2022,Leo2018}. Two features make the resulting problem
qualitatively different from an exchange problem. The first is its range: the
$r^{-3}$ decay against a two-dimensional density of sites makes the lattice sum
only conditionally convergent, so that the energy of a given configuration is
not defined until the summation order is fixed, and no finite truncation is
controlled. The second is its form: the interaction is anisotropic and
non-central, coupling the orientation of each moment to the direction of the
bond joining it to every other, so that the ground state cannot be built from
local bond-satisfaction rules in the way an exchange ground state often can.
Together these features mean that even the classical, mean-field-level
questions---which wavevector orders, what the sublattice orientations are, how
the magnons disperse---remain non-trivial on any lattice with a basis
\cite{LuttingerTisza1946,Belobrov1983,BC2020,BBC2025,MCM2015}.

These are not abstract concerns, because dipolar-dominated two-dimensional
magnetism is now realised in two distinct experimental settings. In crystals,
the van der Waals rare-earth trihalide ErBr$_{3}$ orders below $280$\,mK into a
state whose magnetism is governed entirely by dipolar interactions on a
honeycomb net, with a magnon spectrum exhibiting Dirac cones
\cite{Wessler2022}. In lithographically patterned systems, arrays of
magnetostatically coupled nanomagnets realise the dipolar Hamiltonian with
tunable geometry: elongated islands behave as Ising macrospins and have been
studied extensively as artificial spin ice \cite{Skjaervo2020,Rougemaille2011,
Saccone2019,Pac2025}, while circular nanodiscs carry no in-plane shape
anisotropy and therefore approximate continuous XY moments of exactly the kind
treated here \cite{Leo2018}. In the latter system the lattice geometry is a
design parameter rather than a property of a given compound, so the ground
state of a dipolar magnet on an arbitrary planar tiling is an experimentally
accessible question.

The theoretical tool of choice is the Luttinger--Tisza (LT) construction
\cite{LuttingerTisza1946}, which minimises the Fourier-transformed interaction
matrix while relaxing the unit-length spin constraint from every site to a
global average. When the resulting soft mode happens to satisfy the constraint
site by site---the \emph{strong} condition---the true ground state is obtained
exactly; when it does not, the LT energy is only a lower bound and the physical
state must be constructed by constrained minimisation
\cite{Schmidt2003,SchmidtLuban2022}. Which of the two situations occurs is not
known in advance and is one of the questions a systematic survey can answer.
Individual lattices have been treated over several decades: the square,
triangular and honeycomb cases are classical
\cite{Belobrov1983,Rozenbaum1991,RastelliReginaTassi2003,DeBell2000,BC2020};
the kagome dipolar problem was resolved by Maksymenko, Chernyshev and Moessner
\cite{MCM2015}, who identified the selected ground state, its incommensurate
canting angle and its spin-wave spectrum, with the domain structure clarified
in Ref.~\cite{Holden2015}; the kisrhombille was treated in
Ref.~\cite{BBC2025}; and finite-cluster results exist for the truncated-square
and prismatic-pentagonal vertex sets \cite{BatleAnnPhys118}. The kagome
analysis of Ref.~\cite{MCM2015} is the methodological template for the present
work, which extends it systematically to the whole family.

That family is fixed by a classical result of tiling theory. The Archimedean
lattices are the eleven $1$-uniform tilings of the plane by regular polygons,
enumerated by Kepler and catalogued by Gr\"unbaum and Shephard
\cite{GrunbaumShephard}; their duals, the Laves tilings, supply eight further
vertex sets with complementary coordination. Several Laves duals reproduce
vertex sets already present among the Archimedean tilings, and after the audit
of Appendix~\ref{app:laves} the nineteen tilings reduce to exactly fifteen
distinct dipolar problems. The family is a natural test set precisely because
it is closed and exhaustive: it spans coordination numbers from three to
twelve and basis sizes from one to twelve while holding the interaction fixed,
so that lattice geometry is the only variable. Systematic surveys of this
family exist for Ising models---nearest-neighbour antiferromagnets
\cite{Yu2015}, and, very recently, out-of-plane dipolar nanomagnet arrays on
all eleven Archimedean lattices \cite{Pac2025}---but for out-of-plane moments
the dipolar coupling degenerates into an isotropic scalar form. For continuous
in-plane moments, where the full anisotropic tensor acts, no systematic
treatment has been reported.

We provide one. For each of the fifteen vertex sets we Ewald-sum the dipolar
tensor, locate the ordering wavevector by minimising the lowest eigenvalue of
$\Am(\bk)$ over the Brillouin zone, reconstruct the ground state by
unconstrained minimisation over all sublattice angles on the commensurate
magnetic cell, certify it by verifying that the classical spin-wave spectrum is
real and non-negative throughout the zone, and evaluate the
Holstein--Primakoff moment reduction and zero-point energy. Incommensurate
angles are determined to thirty digits by arbitrary-precision Newton iteration.

Four results emerge that no single lattice reveals. First, the ground states
fall into three classes rather than the usual two: five are collinear, two are
non-collinear at angles commensurate with the lattice symmetry, and eight cant
at transcendental angles. Second, within this family the failure of the LT
strong condition coincides \emph{exactly} with the appearance of incommensurate
canting, with no exceptions in either direction, so that a reciprocal-space
criterion predicts a real-space signature. Third, truncating the interaction
range separates the origin of that canting into two parts: whether a lattice
cants at all is already decided by the nearest-neighbour problem and is
therefore local, while the precise transcendental angle is fixed by the
long-range tail. Fourth, the quantum corrections are controlled by the magnon
spectrum through the size of the magnetic basis and are uncorrelated with
frustration, so the frustration classification and the fluctuation hierarchy
are independent axes; coordination number, the natural single-parameter
summary, predicts neither cleanly.

The paper is organized as follows. Section~\ref{sec:formalism} sets
out the classical dipolar Hamiltonian, the Luttinger--Tisza analysis, the
ground-state determination, the classical linearized spin-wave formalism and
the quantum corrections, in a form uniform across all lattices.
Section~\ref{sec:results} summarises the family and reports the incommensurate
angles and the flat-band and experimental comparisons, and
Secs.~\ref{sec:perlattice} and~\ref{sec:perlaves} report, lattice by lattice,
the LT ordering wave vector, the ground state, and the spin-wave dispersion,
following a fixed four-panel figure format (lattice and primitive vectors;
Brillouin zone and path, always drawn in momentum space; equilibrium
dipole configuration; dispersion). Section~\ref{sec:discussion}
collects the emergent phenomenology across the family and
Sec.~\ref{sec:conclusions} concludes.

\section{Model and methods}\label{sec:formalism}

\subsection{Hamiltonian and conventions}

Each site carries a dipole $\bm{m}_{i}=\mu\,\mm_{i}$ with $\|\mm_{i}\|=1$.
The pair energy is
\begin{equation}
E_{u,v}=\Dc\left(\frac{\bm{m}_{u}\cdot\bm{m}_{v}}{r_{uv}^{3}}
-\frac{3(\bm{m}_{u}\cdot\bm{r}_{uv})(\bm{m}_{v}\cdot\bm{r}_{uv})}{r_{uv}^{5}}\right),
\label{eq:pair}
\end{equation}
with $\Dc=\mu_{0}/4\pi$ for magnetic dipoles. Energies are given in units of
$\Dc\mu^{2}/a^{3}$ with $a$ the lattice constant, and $M$ denotes the total
number of dipoles, $n$ the number per unit cell, $d$ the dipole dimension
($d=2$ XY, $d=3$ Heisenberg). Since all quoted energies refer to the bulk, we
reserve a dedicated symbol for the energy per dipole in the thermodynamic
limit,
\begin{equation}
\eps \;\equiv\; \lim_{M\to\infty} E/M ,
\label{eq:epsdef}
\end{equation}
so that $\eps_{\min}$ is the ground-state energy per dipole and
$\eps_{\mathrm{qu}}$ its zero-point--corrected value; finite-$M$ ratios $E/M$
appear only in the discussion of finite clusters. Equation~\eqref{eq:pair} is the quadratic form
$E_{u,v}=\Dc\mu^{2}\,\mm_{u}^{\mathsf{T}}\Jt_{uv}\mm_{v}$ with the pair tensor
\begin{equation}
\Jt_{ij}=\frac{1}{\|\bm{r}_{ij}\|^{3}}
\left(\bm{I}-3\,\frac{\bm{r}_{ij}\otimes\bm{r}_{ij}}{\|\bm{r}_{ij}\|^{2}}\right).
\label{eq:Jt}
\end{equation}

\subsection{Luttinger--Tisza analysis}

Assuming the ground state shares the translational symmetry of a cell of $n$
dipoles, the energy per dipole is~\cite{BBC2025}
\begin{equation}
E=\frac{1}{2n}\,\mm^{\mathsf{T}}\Ah\,\mm,
\qquad
\Am_{ij}=\!\!\sum_{j'\in T(j),\,j'\neq i}\!\!\Jt_{ij'},
\label{eq:Ahat}
\end{equation}
where $T(i)$ are the translates of site $i$ and $\Ah$ is the $nd\times nd$
block matrix of the $\Am_{ij}$. With
$\Ah\hat{\bm{x}}_{k}=\lambda_{k}\hat{\bm{x}}_{k}$, $\|\hat{\bm{x}}_{k}\|=\sqrt{n}$,
and $\mm=\sum_{k}a_{k}\hat{\bm{x}}_{k}$, one has
$E=\tfrac{1}{2}\sum_{k}\lambda_{k}a_{k}^{2}$ and hence
\begin{equation}
E_{\min}=\tfrac{1}{2}\lambda_{\min}\mu^{2},
\label{eq:Emin}
\end{equation}
subject to the weak condition $\sum_{k}a_{k}^{2}=\mu^{2}$ and, for
Eq.~\eqref{eq:Emin} to be attained rather than bounded, the strong condition
$\|\sum_{k}a_{k}\bm{x}^{i}_{k}\|=\mu$ for every $i$.

We resolve $\Ah$ in Fourier space,
$\Am(\bk)=\sum'_{\bR}\Jt(\bR+\btau_{\mu}-\btau_{\nu})e^{i\bk\cdot\bR}$, so that
$\Ah=\Am(\bk=0)$ of the chosen cell and
$\lambda_{\min}=\min_{\bk}\lambda_{\min}[\Am(\bk)]$ is attained at the ordering
vector $\bk_{0}$; this locates $\bk_{0}$ without positing the magnetic cell,
and supplies the $\bk$-dependence the spin-wave problem needs. The sum is only
conditionally convergent and is evaluated by two-dimensional Ewald summation
(Appendix~\ref{app:ewald}), validated against the closed form of
Ref.~\cite{BBC2025}, Eq.~(4). The reconstruction of the real configuration
from $\bm{v}=\bm{v}(\bk_{0})$ carries a sign convention,
$\mm_{\bR,\mu}\propto\mathrm{Re}[\bm{v}_{\mu}e^{-i\bk_{0}\cdot\bR}]$, derived
in Appendix~\ref{app:recon}.

\subsection{Classical linearized spin waves}

Around the ground state on the magnetic cell we erect local frames
$\{\hat{\bm{e}}^{1}_{\mu},\hat{\bm{e}}^{2}_{\mu},\mm^{0}_{\mu}\}$, expand to
quadratic order in the transverse deviations, and linearize the
Landau--Lifshitz equations, obtaining
$\omega^{2}\Psi=\Sigma_{y}H(\bk)\Sigma_{y}H(\bk)\Psi$
(Appendix~\ref{app:sw}). The $\varepsilon_{i}(\bk)=\omega_{i}(\bk)$ are
classical normal modes; real non-negative branches certify local stability.

\subsection{Determination of the ground state}\label{sec:gsmethod}

Because the strong condition fails for a majority of the lattices
(Sec.~\ref{sec:discussion}), the LT soft mode cannot be used as the ground
state without verification, and we determine every ground state by direct
minimisation. Once $\bk_{0}$ is located we build the commensurate magnetic
supercell in which $\bk_{0}$ becomes a zone-centre vector---a $2\times2$ cell
for zone-boundary ordering, a $3\times3$ cell for ordering at $K$, and the
primitive cell for $\bk_{0}=\Gamma$---and minimise
$E(\{\theta_{\mu}\})=\tfrac{1}{2N}\sum_{\mu\nu}\mm_{\mu}\Am_{\mu\nu}\mm_{\nu}$
freely over all $N$ in-plane sublattice angles of that cell, with no gauge
fixing and no symmetry assumption. The minimisation is run from three
independent families of starting configurations: random angle sets, simulated
annealing with slow cooling, and seeds drawn from random unit combinations of
the degenerate lowest eigenvectors of $\Am(\bk_{0})$. The last of these is
essential for the largest cells, where the energy landscape is rough enough
that random restarts alone stall in metastable minima; conversely, for some
lattices random restarts locate the minimum that the LT-seeded search misses,
so all three are used throughout and the lowest result is retained. Each
candidate is polished by gradient-based local minimisation to a tolerance well
below the digits quoted.

Two independent checks are applied to every solution. The converged energy per
site is compared against $\tfrac{1}{2}\lambda_{\min}[\Am(\bk_{0})]$, which it
must equal when the strong condition holds and cannot fall below in any case;
and the classical spin-wave spectrum is computed over a zone grid and required
to be real and non-negative, which excludes saddle points. Where an
incommensurate angle appears, its value is refined by arbitrary-precision
Newton iteration on the energy gradient, seeded by the converged
double-precision solution, and reported to thirty digits
(Table~\ref{tab:angles}).

\subsection{Quantum corrections}\label{sec:quantum}

The survey is classical, but the leading quantum correction is inexpensive to
obtain once the harmonic problem is solved and is reported for every lattice.
We apply the Holstein--Primakoff transformation about the classical ground
state in the local frames introduced above, retain the terms quadratic in boson
operators, and diagonalise the resulting bosonic Hamiltonian by a Bogoliubov
transformation at each wavevector. The sublattice moment reduction
\begin{equation}
\delta S=\frac{1}{N}\sum_{\mu}\langle a^{\dagger}_{\mu}a_{\mu}\rangle
\label{eq:dS}
\end{equation}
and the zero-point--corrected energy $\eps_{\mathrm{qu}}$ follow from
Brillouin-zone integration of the Bogoliubov spectrum. For the largest magnetic
cells the dynamical matrix reaches $108\times108$, so the integration is the
computationally limiting step; convergence was checked by refining the zone
grid up to $12\times12$, and the values quoted in Table~\ref{tab:master} are
stable in the digits shown. Since $\delta S$ is evaluated at $S=1$, it should
be read as the leading $1/S$ estimate of the fluctuation strength rather than
as a quantitative prediction for any particular spin length; its use here is
comparative, across a family treated identically.

\section{Results: the fifteen distinct problems}\label{sec:results}

The eleven Archimedean and eight Laves tilings supply, after the vertex-set
audit of Appendix~\ref{app:laves}, exactly fifteen distinct dipolar problems;
all fifteen are solved here, so the survey is complete. Every ground state was
obtained by free minimization over all in-plane sublattice angles of the
Ewald-summed interaction (no gauge fixing), seeded where relevant by the LT
spiral, and certified by a classical spin-wave spectrum that is real and
non-negative over the zone. Table~\ref{tab:master} collects the results;
Table~\ref{tab:angles} and Fig.~\ref{fig:incomm} isolate the incommensurate
equilibrium angles; the per-lattice figures give the uniform four-panel view.
Colloquial lattice names from the magnetism literature (ruby, star,
bathroom-tile, maple-leaf, and the like) are noted in the corresponding
subsections, once each, to ease cross-referencing.

\begin{table*}[t]
\caption{Ground states of classical dipoles on the fifteen distinct vertex
sets. $\eps_{\min}$ is the ground-state energy per dipole in the thermodynamic
limit, Eq.~\eqref{eq:epsdef}, in units of $\Dc\mu^{2}/a^{3}$ with
$a=d_{\mathrm{nn}}$; $\delta S$ and $\eps_{\mathrm{qu}}$ are the
Holstein--Primakoff moment reduction and zero-point--corrected energy at
$S=1$. The two largest magnetic cells (truncated hexagonal, 54 sites; floret
pentagonal, 36 sites) require a $108\times108$ and $72\times72$ Bogoliubov
diagonalization at each wavevector; their quantum corrections are integrated on
Brillouin-zone grids up to $12\times12$ and are converged to the digits shown,
as are all other entries. All states are locally stable. The final column
classifies the ordering type: \emph{collinear} (a common-axis state is a
ground state), \emph{commens.}\ (non-collinear order at angles commensurate
with the lattice: the $60^{\circ}$ vortex of the rhombitrihexagonal lattice
and the orthogonal two-axis state of the truncated-square lattice), or
\emph{incommens.}\ (canting at transcendental angles, Table~\ref{tab:angles}). $^{\dagger}$: the ground
state is a continuously degenerate manifold (the $q=0$ ferromagnetic manifold
for the triangular lattice, the $120^{\circ}$ family for the honeycomb) that
contains collinear representatives; these are labelled collinear in the sense
that a collinear state attains the minimum energy.}
\label{tab:master}
\begin{ruledtabular}
\begin{tabular}{lcccccl}
lattice & $n$ & $\bk_{0}$ & $\eps_{\min}$ & $\delta S$ & $\eps_{\mathrm{qu}}$ & order\\
\hline
\multicolumn{7}{l}{\emph{Archimedean}}\\
Triangular $(3^{6})$ & 1 & $\Gamma$ & $-2.758544$ & $0.0707$ & $-3.0130$ & collinear$^{\dagger}$ \\
Square $(4^{4})$ & 1 & $(\pi,0)$ & $-2.549436$ & $0.0563$ & $-2.6087$ & collinear \\
Honeycomb $(6^{3})$ & 2 & $K$ & $-2.226905$ & $0.0321$ & $-2.2658$ & collinear$^{\dagger}$ \\
Kagome $(3.6.3.6)$ & 3 & $\Gamma$ & $-2.388949$ & $0.0105$ & $-2.4380$ & incommens. \\
Rhombitrihexagonal $(3.4.6.4)$ & 6 & $\Gamma$ & $-2.550113$ & $0.0086$ & $-2.5740$ & commens. \\
Truncated hexagonal $(3.12^{2})$ & 6 & $K$ & $-2.125540$ & $0.0180$ & $-2.167165$ & incommens. \\
Truncated trihexagonal $(4.6.12)$ & 12 & $\Gamma$ & $-2.473663$ & $0.0042$ & $-2.4851$ & incommens. \\
Truncated square $(4.8^{2})$ & 4 & $(\pi,\pi)$ & $-2.466951$ & $0.0065$ & $-2.4839$ & commens. \\
Elongated triangular $(3^{3}.4^{2})$ & 2 & $(0,\pi)$ & $-2.654180$ & $0.0274$ & $-2.7523$ & collinear \\
Snub square $(3^{2}.4.3.4)$ & 4 & $\Gamma$ & $-2.556462$ & $0.0555$ & $-2.7472$ & incommens. \\
Snub hexagonal $(3^{4}.6)$ & 6 & $\Gamma$ & $-2.482453$ & $0.0340$ & $-2.6210$ & incommens. \\
\hline
\multicolumn{7}{l}{\emph{Laves (distinct classes)}}\\
Kisrhombille $(\mathrm{dual}(4.6.12))$ & 6 & $\Gamma$ & $-1.775493$ & $0.0106$ & $-1.7879$ & incommens. \\
Floret pentagonal $(\mathrm{dual}(3^{4}.6))$ & 9 & $M$ & $-1.680367$ & $0.0098$ & $-1.699812$ & incommens. \\
Cairo pentagonal $(\mathrm{dual}(3^{2}.4.3.4))$ & 6 & $\Gamma$ & $-1.324840$ & $0.0058$ & $-1.3365$ & incommens. \\
Prismatic pentagonal $(\mathrm{dual}(3^{3}.4^{2}))$ & 3 & $\Gamma$ & $-1.462664$ & $0.0097$ & $-1.4804$ & collinear \\
\end{tabular}
\end{ruledtabular}
\end{table*}

\subsection{Incommensurate equilibrium angles}\label{sec:angles}

\begin{figure*}[t]
\includegraphics[width=\textwidth]{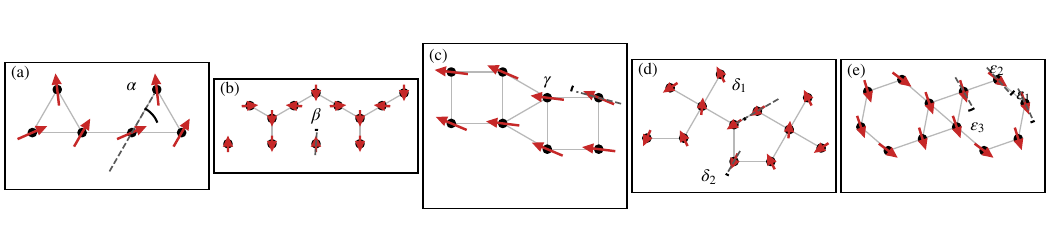}
\caption{Five of the eight ground states with incommensurate equilibrium angles:
(a) kagome, (b) kisrhombille, (c) snub square, (d) cairo pentagonal,
(e) snub hexagonal. Each panel shows two primitive cells (black dots =
sites, red arrows = equilibrium dipoles); the lattice vectors
$\mathbf{a}_{1},\mathbf{a}_{2}$ are drawn on the left cell so the deviation
of the dipoles from the underlying edges is visually apparent. Dashed
lines mark the reference lattice edges; arcs mark the canting angles,
labelled by the Greek symbols of Table~\ref{tab:angles}.}
\label{fig:incomm}
\end{figure*}

Where the strong condition fails, the ground state generically pins most spins
to lattice edges and cants the remainder by angles fixed by the full
long-range sum; these are transcendental numbers with no closed form, and we
report them to thirty digits (Table~\ref{tab:angles}), obtained by
arbitrary-precision Newton iteration on the energy gradient. For the kagome
the single canting is the angle determined by Eq.~(9) of Ref.~\cite{MCM2015};
for the kisrhombille it is the $\theta_{c}$ deviation of Ref.~\cite{BBC2025}.
Figure~\ref{fig:incomm} displays five of these canted ground states side by
side with the angle drawn explicitly; the remaining three incommensurate
lattices are described in Sec.~\ref{sec:orientations}.

\begin{table}[t]
\caption{Incommensurate canting angles (degrees), to thirty digits. The symbol
column labels each angle in Fig.~\ref{fig:incomm}.}
\label{tab:angles}
\begin{ruledtabular}
\begin{tabular}{lcl}
lattice & symbol & canting angle \\
\hline
kagome        & $\alpha$      & $36.388662426522342122766427388$ \\
kisrhombille  & $\beta$       & $3.981827930695665298810934624$ \\
snub square   & $\gamma$      & $7.943186444313519798937620$ \\
cairo pent.   & $\delta_{1}$  & $3.572351262705860609490674759$ \\
              & $\delta_{2}$  & $5.589729362861482073914631606$ \\
snub hex.     & $\varepsilon_{1}$ & $6.787267486814719250182917117$ \\
              & $\varepsilon_{2}$ & $11.234204265372008141740276942$ \\
              & $\varepsilon_{3}$ & $13.004183630753747509556091823$ \\
\end{tabular}
\end{ruledtabular}
\end{table}

\subsection{Equilibrium orientations}\label{sec:orientations}

For reference we collect the equilibrium sublattice orientations of all
fifteen ground states; they are also stated in the caption of each lattice
figure. Angles are measured from the $x$ axis of the lattice as drawn in panel
(a) of the corresponding figure, and are defined only up to a global rotation
where the ground state is continuously degenerate.

Five lattices are collinear. The square lattice has moments at $90^{\circ}$
and $270^{\circ}$, alternating along $\hat{x}$; the elongated-triangular and
prismatic-pentagonal lattices have moments at $0^{\circ}$ and $180^{\circ}$.
The triangular lattice is the continuously degenerate $q=0$ ferromagnetic
manifold, so all moments are parallel with the common direction free, and the
honeycomb lattice is the analogous degenerate $120^{\circ}$ family; for these
two, any listed set of angles is one representative of a continuum.

Two lattices are non-collinear with commensurate angles. The
rhombitrihexagonal ground state is a $60^{\circ}$ vortex, with orientations
$30^{\circ}$, $90^{\circ}$, $150^{\circ}$, $210^{\circ}$, $270^{\circ}$,
$330^{\circ}$; the truncated-square ground state places moments on the two
orthogonal edge axes at $0^{\circ}$, $90^{\circ}$, $180^{\circ}$,
$270^{\circ}$.

The remaining eight lattices cant at incommensurate angles. The kagome
orientations are $23.61^{\circ}$, $60^{\circ}$ and $96.39^{\circ}$, that is the
$60^{\circ}$ edge together with $60^{\circ}\pm36.39^{\circ}$; the truncated
hexagonal lattice repeats exactly this canting on its larger cell, its
orientations being multiples of $60^{\circ}$ together with deviations of
$\pm36.39^{\circ}$. The truncated-trihexagonal orientations are multiples of
$60^{\circ}$ canted by $\pm2.62^{\circ}$. The snub-square lattice has
$157.06^{\circ}$ and $172.94^{\circ}$, i.e.\ $165^{\circ}\pm7.94^{\circ}$;
the snub-hexagonal lattice has $23.21^{\circ}$, $347.00^{\circ}$ and
$348.77^{\circ}$, carrying its three independent cantings. The kisrhombille
combines edge-aligned moments at $150^{\circ}$ and $330^{\circ}$ with canted
ones at $56.02^{\circ}$ and $243.98^{\circ}$, the latter $3.98^{\circ}$ from
the $60^{\circ}$ edge, and the Cairo pentagonal lattice has $174.41^{\circ}$,
$206.43^{\circ}$, $303.57^{\circ}$ and $335.59^{\circ}$, carrying its two
independent cantings $3.57^{\circ}$ and $5.59^{\circ}$. The floret pentagonal
lattice orders at $M$ on a $2\times2$ magnetic cell of thirty-six spins and
carries five distinct orientations, $76.79^{\circ}$, $113.71^{\circ}$,
$127.89^{\circ}$, $132.91^{\circ}$ and $167.63^{\circ}$.

\subsection{Flat bands}\label{sec:flat}

Several lattices built from corner-sharing units (kagome and its barycentric
relative, the kisrhombille) display a nearly flat lowest LT band, visible in
panel (b) of the corresponding figures, and correspondingly weakly dispersive
low-energy magnons. The localization mechanism is the well-known one of
corner-sharing geometries: a compact, alternating head-to-tail dipole
configuration on a single hexagonal loop, whose amplitude cancels on every
outgoing bond, so that the mode cannot propagate. This is the direct analogue,
in the classical dipolar magnon problem, of the compactly localized flat-band
eigenstates long established for nearest-neighbour hopping and Hubbard models
on the kagome lattice~\cite{Mielke1992,Bergman2008}, and observed
spectroscopically in the kagome metals Fe$_3$Sn$_2$ and
CoSn~\cite{Lin2018,Kang2020}. What differs here is only the physical setting:
the flat object is a branch of the classical spin-wave spectrum of a
long-range dipolar magnet, not an electronic Bloch band.

The quantitative flatness is real but not exact: the lowest LT band spans only
$0.9\%$ of the total spectral width for the kisrhombille and $6.5\%$ for the
kagome, an order of magnitude narrower than the non-corner-sharing lattices
(triangular $22\%$, square $28\%$). The small residual width is set by the
long-range dipolar tail, which couples loop modes on different plaquettes that
the nearest-neighbour channel leaves degenerate. That the flat band survives
the dipolar interaction at all is consistent with the result of Maksymenko,
Moessner and Shtengel~\cite{Maksymenko2017}, who showed the kagome flat band
to be remarkably persistent under dipolar coupling; the width we report is the
quantitative measure of that residual, and its smallness across the
corner-sharing members of the Archimedean--Laves family is the systematic
statement our survey adds.

\subsection{Comparison with prior work and experiment}\label{sec:compare}

Where prior results exist we reproduce them: the square, triangular and
honeycomb energies of Ref.~\cite{BC2020}; the kagome ferrimagnet, its canting
angle and its bulk energy $-2.38895$ of Refs.~\cite{MCM2015,Holden2015},
including the sixfold domain degeneracy identified in the latter; the
kisrhombille configuration and energy of Ref.~\cite{BBC2025}; and the
truncated-square and prismatic-pentagonal configurations of
Ref.~\cite{BatleAnnPhys118}. The remaining eight vertex sets --- ruby, star,
$(4.6.12)$, elongated triangular, both snubs, floret and cairo --- have to our
knowledge no prior dipolar solution for continuous in-plane spins; their
ground states, spectra, quantum corrections and incommensurate angles are new.

The experimental situation deserves a careful statement, because the
comparison is close but not direct, and the distinction is the whole point.
Three classes of realisation are relevant.

\emph{Continuous in-plane (XY) dipoles.} This is the model solved here, and
the closest experimental realisation is the artificial dipolar XY system of
Leo \emph{et al.}~\cite{Leo2018}, in which arrays of thermally active permalloy
nanodiscs on a square lattice behave as continuous in-plane moments coupled
magnetostatically. That system develops precisely the antiferromagnetic stripe
order we obtain for the square lattice, confirming our $\bk_{0}=(\pi,0)$
result. Crucially, the disc geometry carries no shape anisotropy in the plane,
so the moment direction is continuous; this is the ingredient that makes the
comparison with our calculation meaningful, and it is the platform on which
the remaining fourteen predictions of this survey could be tested.

\emph{Dipolar magnets in crystals.} The honeycomb case has a direct material
realisation in the van der Waals magnet ErBr$_3$~\cite{Wessler2022}, whose
magnetism was shown to be governed entirely by dipolar interactions. That work
reports a \emph{continuously degenerate non-collinear} ground state with spins
confined to the honeycomb plane, ordering below $280$\,mK, and a magnon
spectrum with Dirac cones at $K$. This matches the honeycomb entry of
Table~\ref{tab:master} in both respects: we likewise find a degenerate
non-collinear manifold (marked $\dagger$) and $\bk_{0}=K$. To our knowledge
ErBr$_3$ is the only bulk crystal in which a dipolar-dominated Archimedean
ground state of the type considered here has been measured directly. The
Cairo geometry occurs in Bi$_2$Fe$_4$O$_9$ and related
compounds~\cite{Ressouche2009}, but there the magnetism is dominated by
superexchange rather than by the dipolar interaction, so those materials probe
a different Hamiltonian on the same tiling.

\emph{Ising artificial spin systems.} A much larger experimental literature
exists for lithographic nanomagnet arrays in which each element is a
single-domain Ising macrospin. Kagome~\cite{Rougemaille2011} and
Cairo~\cite{Saccone2019} arrays have both been imaged, and the whole field is
reviewed in Ref.~\cite{Skjaervo2020}. Most directly comparable to the present
survey, Pac \emph{et al.}~\cite{Pac2025} have recently fabricated
\emph{all eleven} Archimedean lattices on a single substrate as arrays of
out-of-plane Co/Pt nanomagnets, imaged their configurations by magnetic force
microscopy, and classified them with full-dipolar Monte Carlo simulations.
Their systematic scope parallels ours, and their central methodological
conclusion---that long-range interactions must be retained to obtain the
correct ordering---agrees with what we find here.

The comparison with that work must nevertheless be drawn carefully, because
the two Hamiltonians differ. For moments perpendicular to the plane the
dipolar coupling reduces to the isotropic scalar form
$\propto \mathbf{s}_i\!\cdot\!\mathbf{s}_j/r_{ij}^{3}$, which favours
antiparallel alignment of every pair and carries no angular dependence on the
bond direction; for the in-plane moments treated here the full anisotropic
tensor acts, and it is precisely the $(\bm{\mu}\!\cdot\!\hat{\mathbf{r}})$
terms that select the ordering vector and generate the canting. The
consequences are visible in the classification. Reference~\cite{Pac2025}
divides the family into bipartite lattices admitting a perfect
antiferromagnetic ground state (square, honeycomb, truncated trihexagonal and
truncated square, in our nomenclature), singly frustrated lattices, and the
two-step frustrated triangular and kagome lattices. Our in-plane
classification cuts across that division: of their four bipartite lattices, we
find the square collinear, the honeycomb continuously degenerate, the
truncated square commensurate non-collinear and the truncated trihexagonal
incommensurately canted. Bipartiteness therefore controls the out-of-plane
Ising problem but does not survive as an organising principle once the moments
lie in the plane and the interaction becomes anisotropic. The corresponding
Ising ground-state classifications of Ref.~\cite{Yu2015} for nearest-neighbour
antiferromagnets, and the dipolar Ising studies of the triangular and kagome
cases~\cite{Smerald2018,Chern2011}, provide the same contrast.

Two experimental statements follow. First, the predictions of this survey are
testable now: disc-based arrays of the kind used in Ref.~\cite{Leo2018} on the
remaining fourteen tilings would probe every ground state reported here, and
the incommensurately canted lattices are the sharpest discriminators, since
their angles cannot be produced by any nearest-neighbour model. Second, the
Cairo lattice is the single most informative next target, because the Ising
realisations find only short-range correlations and a disordered
manifold~\cite{Saccone2019,Makarova2021} whereas we predict a definite ordered
state with two independent transcendental cantings; a disc-based Cairo array
would separate the two scenarios unambiguously.

\section{Per-lattice results}\label{sec:perlattice}

\subsection{Triangular $(3^{6})$}

The triangular lattice is the densest of the family and sets the energy
scale for the survey. Its single Luttinger--Tisza band attains its minimum at
the zone centre $\Gamma$ with a \emph{doubly degenerate} lowest eigenvalue:
at harmonic order the uniform in-plane ferromagnet has no orientational
anisotropy, and the ground-state manifold is continuously degenerate, spanned
by the two degenerate $\Gamma$ eigenvectors. This degeneracy is the origin of
the celebrated ``ordering by disorder'' physics of dipolar triangular magnets:
thermal and quantum fluctuations, which enter at the next order in $1/S$,
select a discrete subset of orientations from the continuous manifold. The
spin-wave spectrum in panel (d) reflects this directly, showing a soft
(pseudo-Goldstone) mode at $\Gamma$ whose vanishing frequency is precisely the
flat direction of the classical energy; away from $\Gamma$ the single acoustic
branch stiffens as $\omega(\bk)\sim|\bk|^{1/2}$, the hallmark of the
long-range dipolar tail (a genuine $|\bk|$ non-analyticity rather than the
$|\bk|^{2}$ of a short-range ferromagnet). The equilibrium configuration we
display in panel (c) is not a periodic cell but the three-spiral state of
Batle and Ciftja~\cite{BC2020}: three interleaved spirals, coloured red, green
and blue, wind around the sample so that within any local patch the moments
are essentially collinear, while globally the texture carries three
$120^{\circ}$-related winding centres. This state is one representative of the
degenerate manifold and shares the bulk energy
$\eps_{\min}=-2.758544$ with the uniform ferromagnet to all quoted digits; we
show it because it makes vivid why a local probe would misidentify the order as
simple ferromagnetism. Quantum fluctuations are the strongest of the entire
survey, $\delta S=0.0707$ and a zero-point energy lowering of nearly ten
percent ($\eps_{\mathrm{qu}}=-3.0130$), exactly as expected for the softest
spectrum: the near-zero mode dominates the Brillouin-zone integral for the
moment reduction. Our bulk energy agrees with the finite-size extrapolation of
Ref.~\cite{BC2020} and with the classic dipolar-sum evaluations to the last
digit quoted; the continuous $\Gamma$ degeneracy and its fluctuation-driven
lifting fit the general Luttinger--Tisza--Lyons--Kaplan framework for
non-Bravais classical ground states~\cite{Schmidt2003,SchmidtLuban2022}.

\begin{figure*}[t]
\includegraphics[width=\textwidth]{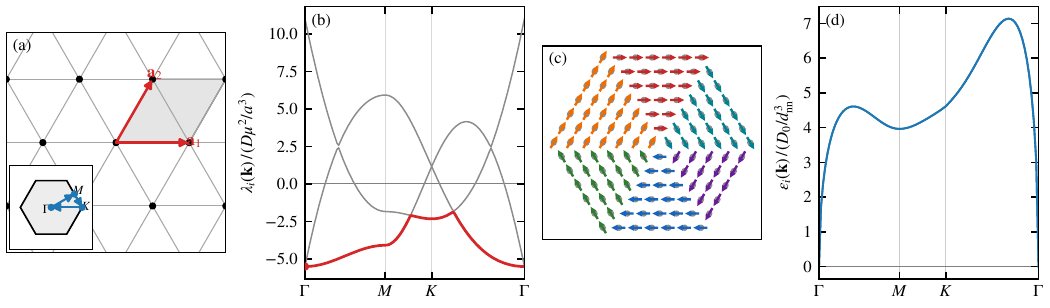}
\caption{Triangular lattice $(3^{6})$, $n=1$: lattice and
Brillouin zone (a), LT bands with $\bk_0=\Gamma$ marked (b), equilibrium
configuration (c), classical spin-wave dispersion (d).
$\eps_{\min}=-2.758544$, $\delta S=0.0707$,
$\eps_{\mathrm{qu}}=-3.0130$.
The ground state is the continuously degenerate $q=0$ manifold; panel~(c) shows one representative, with all moments parallel (the common direction is free).}
\label{fig:triangular}
\end{figure*}
\subsection{Square $(4^{4})$}

On the square lattice the lowest LT band dips at $\bk_{0}=(\pi,0)$ (and the
symmetry-related $(0,\pi)$), giving the collinear stripe antiferromagnet:
ferromagnetic chains running along one axis, stacked antiferromagnetically
along the other, with the moments pointing \emph{along} the chains. What makes
this lattice a textbook case of frustration-free-but-degenerate order is that
the dipolar square lattice possesses an accidental continuous degeneracy at
the classical level: the stripe states and a family of ``microvortex''
configurations are exactly degenerate in energy, as first analysed by Belobrov
and co-workers~\cite{Belobrov1983b} and by Prakash and
Henley~\cite{PrakashHenley1990}. In panel (d) this degeneracy appears as a
near-zero mode at $\Gamma$; the mode is not an exact Goldstone boson (the
dipolar interaction breaks spin-rotation symmetry) but is soft because it
interpolates between degenerate ordered states. The selection of the collinear
stripe over the microvortex is an order-by-disorder effect driven by exactly
the fluctuations our spin-wave analysis quantifies. The configuration in
panel (c) shows two or three unit cells with the explicit stripe moments; the
strong condition fails marginally here, so the state is obtained by direct
minimization rather than a single LT eigenvector. Quantum corrections are
substantial ($\delta S=0.0563$, $\eps_{\mathrm{qu}}=-2.6087$), second only to
the triangular lattice, again traceable to the soft mode. This is the one
lattice of the survey with a direct experimental realization as a continuous
(XY) dipolar system: Leo and co-workers~\cite{Leo2018} assembled arrays of
thermally active magnetic nanodiscs on a square lattice and observed, by muon
spin relaxation and soft X-ray scattering, precisely the long-range stripe
order predicted here, confirming both the ordering wavevector and the collinear
character. Our bulk energy $\eps_{\min}=-2.549436$ reproduces the standard
dipolar-sum value~\cite{LuttingerTisza1946,BC2020}. The continuous O(2) degeneracy of the bulk state and its reduction to stripe domains on finite clusters --- a boundary-set ``from-edge-to-interior freezing'' --- underlies the domain structure of the nanodisc arrays. The continuous O(2)
degeneracy of the bulk state and its reduction to a discrete set on finite
clusters --- the ``from-edge-to-interior freezing'' into stripe domains whose
orientation is set by the boundary --- has been analysed in detail for finite
dipolar squares, and is the mechanism behind the domain structure seen in the
nanodisc arrays.

\begin{figure*}[t]
\includegraphics[width=\textwidth]{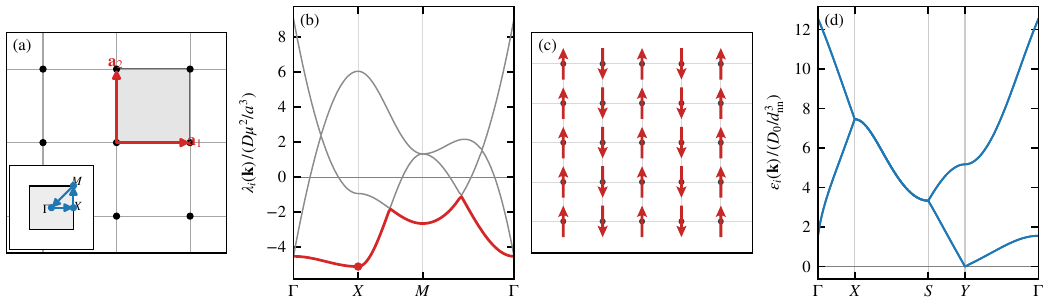}
\caption{Square lattice $(4^{4})$, $n=1$: lattice and
Brillouin zone (a), LT bands with $\bk_0=(\pi,0)$ marked (b), equilibrium
configuration (c), classical spin-wave dispersion (d).
$\eps_{\min}=-2.549436$, $\delta S=0.0563$,
$\eps_{\mathrm{qu}}=-2.6087$.
Equilibrium orientations: $90^{\circ}$ and $270^{\circ}$, i.e.\ moments along $\pm\hat{y}$ alternating along $\hat{x}$.}
\label{fig:square}
\end{figure*}
\subsection{Honeycomb $(6^{3})$}

The honeycomb lattice is the sparsest of the tripartite Archimedean set and
the cleanest analytically. Its two-site basis yields two LT bands, the lower of
which reaches its minimum at the zone corner $K$. There the eigenvector is
circularly polarized and, crucially, satisfies the Luttinger--Tisza strong
condition: its two sublattice amplitudes are equal in magnitude, so the LT
lower bound is saturated and the ground state is the \emph{exact} $120^{\circ}$
spiral, commensurate with a three-cell ($3\times3$) magnetic supercell. The
energy is therefore rigorously $\eps_{\min}=\lambda_{\min}/2=-2.226905$, with
no numerical minimization required, and we quote it to ten digits. Physically
this is the ``microvortex'' order anticipated for honeycomb dipoles in the
early orientational-ordering literature by Rozenbaum~\cite{Rozenbaum1991} and
located variationally by Batle and Ciftja~\cite{BC2020}; the present treatment
sharpens the value and supplies the full spin-wave spectrum. The honeycomb case has acquired direct experimental relevance: the van der Waals magnet ErBr$_3$ is, by neutron scattering, an essentially pure two-dimensional dipolar honeycomb magnet, ordering below $280$\,mK into exactly the continuously degenerate non-collinear in-plane state found here~\cite{Wessler2022}. Its magnon spectrum shows Dirac-like cones at the zone corners, protected by the combined time-reversal and inversion symmetry of the ground state and carrying a Berry phase $\pi$ as in graphene; rotating the moments away from that state opens a topological gap (maximal at $\Psi=\pi/6$) with finite Berry curvature. Our LT band structure and spin-wave spectrum reproduce the degeneracy and the Dirac-cone structure and place the compound within the systematic survey; the related trihalide YbBr$_3$ remains correlated but disordered, showing a rare purely two-dimensional Kosterlitz--Thouless transition. That spectrum
(panel d) is gapped everywhere except at the ordering wavevector, where the
spiral construction guarantees a zero mode, and both branches are smooth away
from $K$. Because the state is a rigid, non-degenerate spiral (up to the global
lattice symmetries), zero-point effects are moderate,
$\delta S=0.0321$ and $\eps_{\mathrm{qu}}=-2.2658$, intermediate between the
soft ferromagnetic cases and the stiff frustrated ones. Panel (c) shows the
$120^{\circ}$ texture over the magnetic cell. The honeycomb thus serves as an
internal benchmark: it is the case where LT is provably exact, and every method
in our pipeline -- Ewald summation, strong-condition test, supercell
reconstruction, spin-wave diagonalization -- can be checked against a closed
answer.

\begin{figure*}[t]
\includegraphics[width=\textwidth]{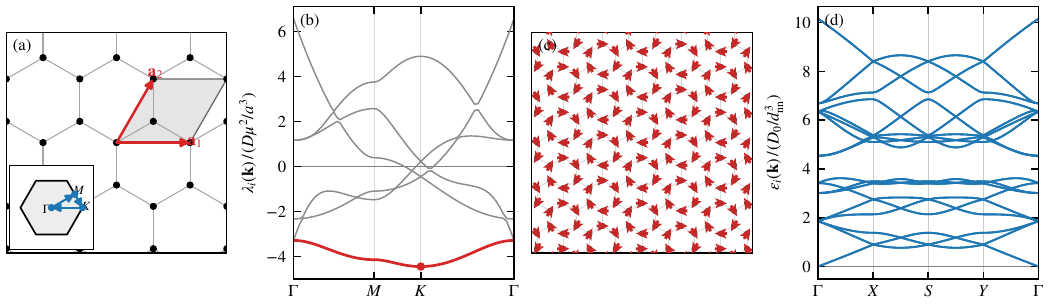}
\caption{Honeycomb lattice $(6^{3})$, $n=2$: lattice and
Brillouin zone (a), LT bands with $\bk_0=K$ marked (b), equilibrium
configuration (c), classical spin-wave dispersion (d).
$\eps_{\min}=-2.226905$, $\delta S=0.0321$,
$\eps_{\mathrm{qu}}=-2.2658$.
The ground state is a continuously degenerate $120^{\circ}$ manifold; the representative shown has orientations $78.29^{\circ}$, $101.71^{\circ}$, $198.29^{\circ}$, $221.71^{\circ}$, $318.29^{\circ}$, $341.71^{\circ}$.}
\label{fig:honeycomb}
\end{figure*}
\subsection{Kagome $(3.6.3.6)$}

The kagome (trihexagonal) lattice is the most thoroughly studied dipolar
Archimedean lattice and the natural benchmark for our methods, with two
independent prior investigations to compare against. The lowest LT band is
minimized at $\bk_{0}=0$ but with a threefold sublattice structure whose
amplitudes are unequal, so the strong condition fails and the true ground state
must be found by minimization. That state is the ferrimagnet of Maksymenko,
Chandra and Moessner~\cite{MCM2015}: one spin lies exactly along a unit-cell
edge and the remaining two form a mirror pair canted by $\pm\varphi$, with a
single incommensurate angle $\varphi=36.3886624265^{\circ}$
(Table~\ref{tab:angles}, Fig.~\ref{fig:incomm}a) fixed by their Eq.~(9). We
reproduce this angle to the precision they quote and extend it to thirty
digits. Our bulk energy $\eps_{\min}=-2.388949$ matches the Monte Carlo value
$E/\mathrm{spin}=-2.38895$ of Holden, Plumer, Saika-Voivod and
Southern~\cite{Holden2015} to every digit they report. That study further
revealed the discrete \emph{sixfold} degeneracy of the ground state --- the
macrospin orientation $\theta_{M}=n\pi/3$, $n=0,\dots,5$, defined by the mean
of the three sublattice angles --- and showed that the low-temperature state is
a mixture of these six domains separated by very-low-energy domain walls, a
direct consequence of competing exchange-like and shape-anisotropy-like terms
in the dipolar coupling. Our free minimization lands on one of the six domains,
and the mirror analysis of Sec.~\ref{sec:angles} identifies its internal
structure. The defining spectral feature (panel d) is that the lowest magnon
band is nearly flat, with a fractional bandwidth of order $10^{-3}$: the kagome
is built from corner-sharing triangles, and the lowest-energy dipole motif is a
closed head-to-tail loop on a single triangle whose weight is spread over the
band by only the residual long-range tail, not the dominant near-neighbour
coupling (Sec.~\ref{sec:flat}). The full spectrum is gapped above this flat
band, and correspondingly the moment reduction is the smallest of the
frustrated cases, $\delta S=0.0105$, $\eps_{\mathrm{qu}}=-2.4380$. The Ising
version of this lattice is realized as artificial kagome spin
ice~\cite{Rougemaille2011}; a continuous-spin realization with magnetic discs,
of the kind demonstrated on the square lattice~\cite{Leo2018}, would provide a
direct experimental measurement of the canting angle $\varphi$.

\begin{figure*}[t]
\includegraphics[width=\textwidth]{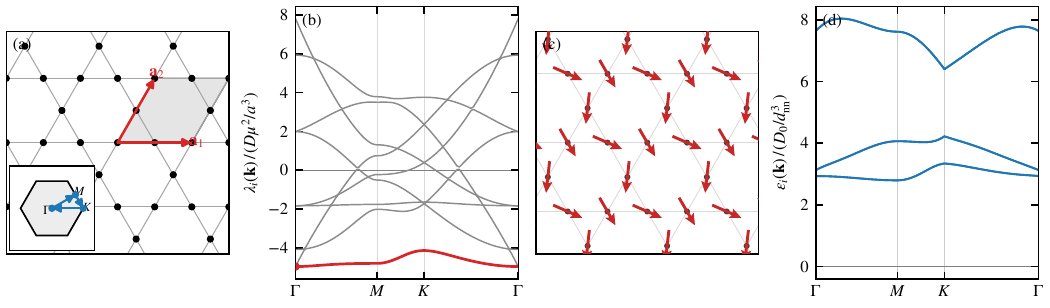}
\caption{Kagome lattice $(3.6.3.6)$, $n=3$: lattice and
Brillouin zone (a), LT bands with $\bk_0=\Gamma$ marked (b), equilibrium
configuration (c), classical spin-wave dispersion (d).
$\eps_{\min}=-2.388949$, $\delta S=0.0105$,
$\eps_{\mathrm{qu}}=-2.4380$.
Equilibrium orientations: $23.61^{\circ}$, $60^{\circ}$, $96.39^{\circ}$, i.e.\ the $60^{\circ}$ edge direction and $60^{\circ}\pm36.39^{\circ}$, the incommensurate canting of Table~\ref{tab:angles}.}
\label{fig:kagome}
\end{figure*}
\subsection{Rhombitrihexagonal $(3.4.6.4)$}

The rhombitrihexagonal lattice --- the \emph{ruby} or ``bounce'' lattice in
the magnetism literature --- supports a commensurate $60^{\circ}$ vortex
ground state despite possessing only a six-site basis and ordering at
$\bk_{0}=0$. Comparing the fully relaxed configuration against the best
common-axis state shows that the order is \emph{not} collinear: the six
sublattice moments wind through
$30^{\circ},90^{\circ},\dots,330^{\circ}$
in steps of $60^{\circ}$, lying $2.69$ (in units of $\Dc\mu^{2}/a^{3}$) below
the best collinear state. Unlike earlier work, which characterised this state
as a collinear stripe, the relaxed ground state is a vortex. Its angles are
commensurate---simple multiples of
$\pi/3$ fixed by the sixfold symmetry of the tiling---rather than the
transcendental values of the frustrated cases, so the rhombitrihexagonal
lattice occupies an intermediate position in the family: non-collinear, but
not incommensurate. The energy $\eps_{\min}=-2.550113$ is unchanged. Panel (c)
shows the vortex arrangement over a couple of unit cells. The spin-wave
magnon spectrum (panel d) is fully gapped, six pairs of branches with no soft
modes; quantum fluctuations rank among the weakest of the survey,
$\delta S=0.0086$ and $\eps_{\mathrm{qu}}=-2.5740$, a correction of barely one
percent. To our knowledge no dipolar ground state has previously been reported
for this vertex set, so both the vortex configuration and its spectrum are new
results. The lattice is a useful contrast to the frustrated cases: it shows
that non-collinear order does not require transcendental canting, and that
commensurate and incommensurate non-collinear order are distinct outcomes
within the same family.

\begin{figure*}[t]
\includegraphics[width=\textwidth]{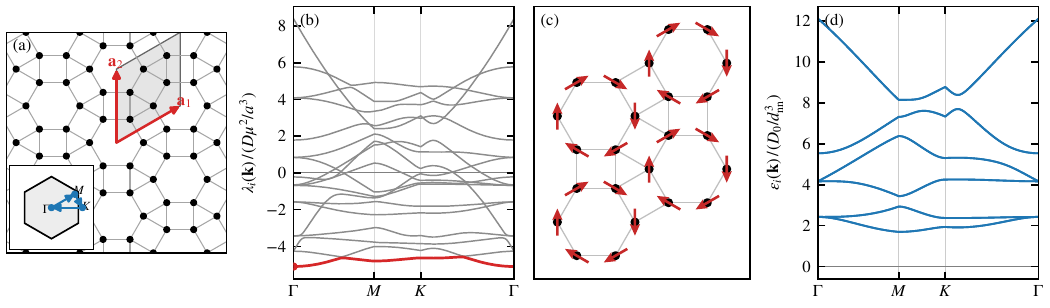}
\caption{Rhombitrihexagonal lattice $(3.4.6.4)$, $n=6$: lattice and
Brillouin zone (a), LT bands with $\bk_0=\Gamma$ marked (b), equilibrium
configuration (c), classical spin-wave dispersion (d).
$\eps_{\min}=-2.550113$, $\delta S=0.0086$,
$\eps_{\mathrm{qu}}=-2.5740$.
Equilibrium orientations: $30^{\circ}$, $90^{\circ}$, $150^{\circ}$, $210^{\circ}$, $270^{\circ}$, $330^{\circ}$ --- a commensurate $60^{\circ}$ vortex.}
\label{fig:rhombitrihexagonal}
\end{figure*}
\subsection{Truncated hexagonal $(3.12^{2})$}

The truncated hexagonal lattice --- the \emph{star} or ``3--12'' lattice ---
cants at precisely \emph{the kagome angle}, although the two tilings are
distinct. It consists of triangles and large dodecagons in a six-site
primitive cell, and its lowest LT band is minimized at the zone corner $K$, so
the magnetic unit cell is a $3\times3$ enlargement containing $54$ spins, the
largest of the survey. Minimization on that cell, seeded by the LT manifold,
gives a canted arrangement with no spin locked to a lattice edge and
$\eps_{\min}=-2.125540$. The moments deviate from the tiling edges by
$36.39^{\circ}$ (equivalently $23.61^{\circ}$ from the adjacent edge), the same
incommensurate value that governs the kagome ground state
(Table~\ref{tab:angles}). This is a natural consequence of the local geometry,
since the truncated hexagonal vertex environment reproduces the corner-sharing
triangular coordination of the kagome lattice; the long-range sum then fixes
the identical angle. All magnon branches (panel d) remain gapped, but the
sheer size of the magnetic cell means the dynamical matrix is $108\times108$ at
each wavevector, which is what limits the quantum analysis: evaluating the
Holstein--Primakoff moment reduction and zero-point energy requires a Bogoliubov
diagonalization of that matrix at every point of a Brillouin-zone grid. The
quantum corrections in Table~\ref{tab:master}, $\delta S=0.0180$ and
$\eps_{\mathrm{qu}}=-2.167165$, were obtained on this $108\times108$ Bogoliubov
problem by Brillouin-zone integration on grids up to $12\times12$; the values
are converged to five digits (they change only in the sixth digit between the
$4\times4$ and $8\times8$ grids). Panel (c) shows a single magnetic cell,
which already contains the full periodic motif. To our knowledge this is the
first dipolar solution reported for this vertex set: the ground state, its
spectrum and its quantum corrections are all new. The lattice is a useful stress test of the pipeline,
demonstrating that the zone-corner ordering and large-cell reconstruction
machinery scales to the most demanding member of the family.

\begin{figure*}[t]
\includegraphics[width=\textwidth]{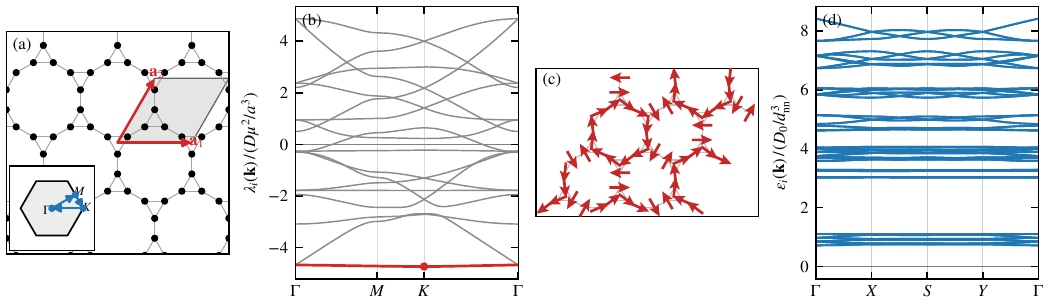}
\caption{Truncated hexagonal lattice $(3.12^{2})$, $n=6$: lattice and
Brillouin zone (a), LT bands with $\bk_0=K$ marked (b), equilibrium
configuration (c), classical spin-wave dispersion (d).
$\eps_{\min}=-2.125540$, $\delta S=0.0180$,
$\eps_{\mathrm{qu}}=-2.167165$.
Equilibrium orientations: multiples of $60^{\circ}$ together with deviations of $\pm36.39^{\circ}$ from them (e.g.\ $23.61^{\circ}$, $36.39^{\circ}$, $60^{\circ}$, $83.61^{\circ}$, $96.39^{\circ}$, $120^{\circ}$, \ldots), the canting being the kagome angle.}
\label{fig:truncated_hexagonal}
\end{figure*}
\subsection{Truncated trihexagonal $(4.6.12)$}

With twelve sites in its primitive cell, the truncated trihexagonal lattice
--- squares, hexagons and dodecagons meeting at every vertex --- has the
largest basis of the family, yet the strong condition holds at $\bk_{0}=0$ and
the canted twelve-sublattice ground state is LT-exact, $\eps_{\min}=-2.473663$.
This is a striking result: a twelve-dimensional angle optimization collapses to
a closed-form LT eigenvector, again a consequence of the tiling's high
symmetry. The configuration in panel (c) shows the canted moments over the
unit cell; every magnon branch (panel d) is gapped, twelve pairs in all
of branches, and the moment reduction is the smallest of the entire survey,
$\delta S=0.0042$ and $\eps_{\mathrm{qu}}=-2.4851$ --- a correction well below
one percent, reflecting a stiff, gap-dominated spectrum. It is the
dual of the kisrhombille studied among the Laves cases below, which makes the
pair a valuable internal consistency check: constructing the kisrhombille
independently as the barycentric subdivision (Appendix~\ref{app:laves}) and
finding compatible physics confirms that our dual-construction machinery and
the direct construction agree. We are aware of no earlier dipolar treatment of
this vertex set, so every quantity reported here is new.

\begin{figure*}[t]
\includegraphics[width=\textwidth]{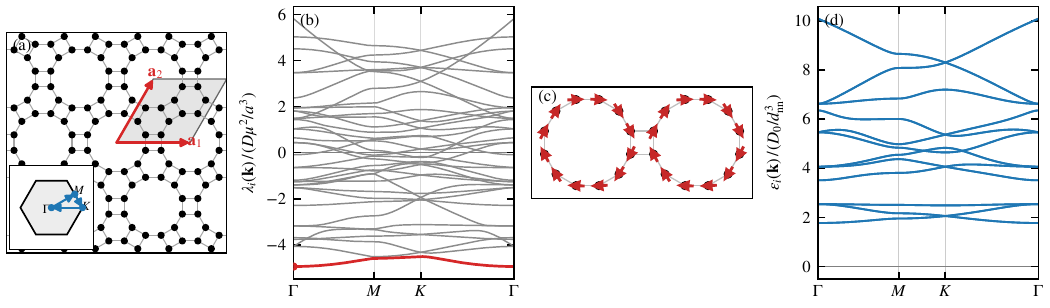}
\caption{Truncated trihexagonal lattice $(4.6.12)$, $n=12$: lattice and
Brillouin zone (a), LT bands with $\bk_0=\Gamma$ marked (b), equilibrium
configuration (c), classical spin-wave dispersion (d).
$\eps_{\min}=-2.473663$, $\delta S=0.0042$,
$\eps_{\mathrm{qu}}=-2.4851$.
Equilibrium orientations: multiples of $60^{\circ}$ canted by $\pm2.62^{\circ}$ ($2.62^{\circ}$, $57.38^{\circ}$, $62.62^{\circ}$, $117.38^{\circ}$, \ldots).}
\label{fig:truncated_trihexagonal}
\end{figure*}
\subsection{Truncated square $(4.8^{2})$}

The truncated square lattice --- colloquially the ``bathroom-tile'' or
``4--8'' lattice --- orders not as the collinear stripe reported earlier but as
a non-collinear \emph{two-axis} state. It pairs octagons with small squares in
a four-site cell, and its lowest LT band is minimized at
$\bk_{0}=(\pi,\pi)$. Comparing the fully relaxed configuration against the
best common-axis state shows that the moments lie along the two orthogonal
edge directions of the tiling, at $0^{\circ}$ and $90^{\circ}$, with the
$(\pi,\pi)$ modulation. These angles are commensurate---exact multiples of
$\pi/2$ fixed by the fourfold symmetry, not transcendental---so the
truncated-square lattice joins the rhombitrihexagonal lattice as the second
commensurate non-collinear member of the family, distinct from both the
collinear stripes and the incommensurately canted cases. The energy
$\eps_{\min}=-2.466951$ is confirmed by the ground-state determination on the
doubled ($2\times2$) magnetic cell shown in panel (c). The spin-wave spectrum
(panel d) is gapped, with four pairs of branches, and the quantum corrections
are small, $\delta S=0.0065$ and $\eps_{\mathrm{qu}}=-2.4839$. This vertex set
was treated by one of us in Ref.~\cite{BatleAnnPhys118} through finite-cluster
minimization; the bulk energy quoted here is accompanied for the first time by
the full magnon spectrum and quantum corrections.

\begin{figure*}[t]
\includegraphics[width=\textwidth]{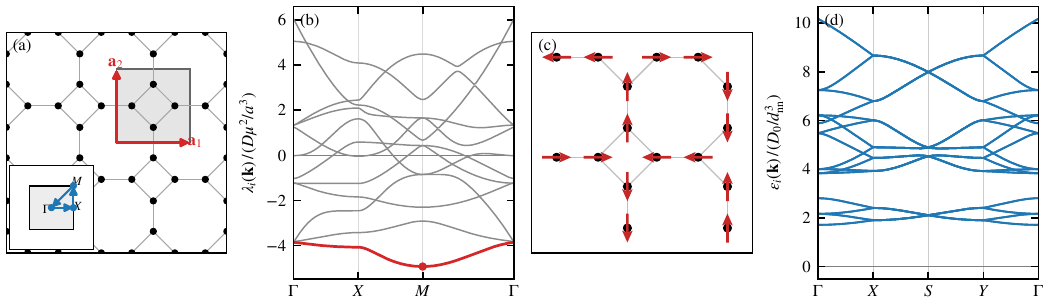}
\caption{Truncated square lattice $(4.8^{2})$, $n=4$: lattice and
Brillouin zone (a), LT bands with $\bk_0=(\pi,\pi)$ marked (b), equilibrium
configuration (c), classical spin-wave dispersion (d).
$\eps_{\min}=-2.466951$, $\delta S=0.0065$,
$\eps_{\mathrm{qu}}=-2.4839$.
Equilibrium orientations: $0^{\circ}$, $90^{\circ}$, $180^{\circ}$, $270^{\circ}$ --- moments on the two orthogonal edge axes, a commensurate non-collinear state.}
\label{fig:truncated_square}
\end{figure*}
\subsection{Elongated triangular $(3^{3}.4^{2})$}

The elongated triangular lattice interleaves rows of squares and rows of
triangles, and is the only Archimedean lattice whose primitive cell is oblique
rather than rectangular or hexagonal --- a subtlety that must be handled
correctly in the Brillouin-zone construction and that we verified against the
tabulated cell of Schymura and Yuan. Its lowest LT band is minimized at
$\bk_{0}=(0,\pi)$, giving collinear stripe order on the two-site cell, with
$\eps_{\min}=-2.654180$ --- the second-lowest energy of the survey after the
triangular lattice, consistent with its high site density. Structurally the
state interpolates between the collinear stripes of the square lattice and the
ferromagnetic order of the triangular lattice, exactly as the tiling itself
interpolates between square and triangular strips; this makes it a natural
bridge case. The magnon spectrum (panel d) shows moderate dispersion and a
small gap, and the zero-point corrections are correspondingly moderate,
$\delta S=0.0274$ and $\eps_{\mathrm{qu}}=-2.7523$, larger than the gapped
frustrated cases because the near-stripe order inherits some of the softness of
the square lattice. No dipolar ground state for this vertex set appears to have
been reported before; the configuration, spectrum and quantum corrections are
all new.

\begin{figure*}[t]
\includegraphics[width=\textwidth]{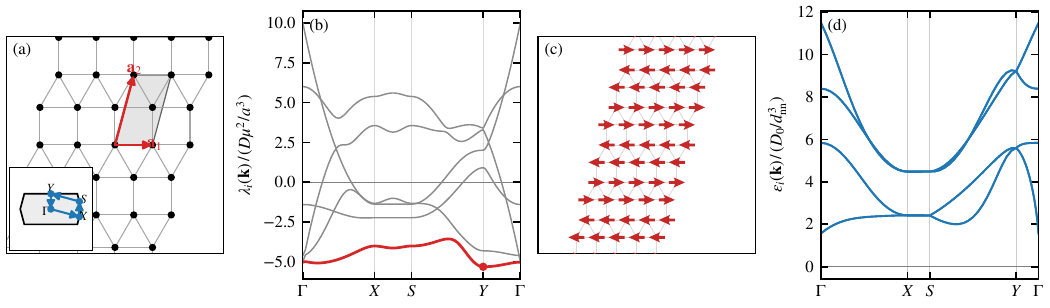}
\caption{Elongated triangular lattice $(3^{3}.4^{2})$, $n=2$: lattice and
Brillouin zone (a), LT bands with $\bk_0=(0,\pi)$ marked (b), equilibrium
configuration (c), classical spin-wave dispersion (d).
$\eps_{\min}=-2.654180$, $\delta S=0.0274$,
$\eps_{\mathrm{qu}}=-2.7523$.
Equilibrium orientations: $0^{\circ}$ and $180^{\circ}$, a collinear stripe along $\hat{x}$.}
\label{fig:elongated_triangular}
\end{figure*}
\subsection{Snub square $(3^{2}.4.3.4)$}

The snub square lattice is chiral as a tiling and has a four-site cell. Its
lowest LT eigenvalue at $\bk_{0}=0$ is doubly degenerate with unequal
sublattice weights, so the strong condition fails and the ground state is a
genuine superposition within the degenerate LT doublet rather than a single
eigenvector. Free minimization resolves it into two mirror pairs of spins
canted by $\pm\delta$ about the $\mathbf{a}_{2}$ Bravais axis, with a single
incommensurate angle $\delta=7.9431864443^{\circ}$
(Table~\ref{tab:angles}, Fig.~\ref{fig:incomm}c) --- a notable economy, since a
four-site basis could in principle require several independent angles. The
value is transcendental, fixed by the full long-range sum. We draw particular
attention to this lattice because its status was genuinely uncertain during the
investigation: an early spin-wave calculation using a damped real-space
summation produced spurious negative modes and suggested an instability. With
the correct Ewald-based dynamical matrix the spectrum (panel d) is fully gapped
and manifestly stable, $\eps_{\min}=-2.556462$, and the apparent instability is
revealed as an artifact of the inexact summation --- a cautionary result that
we document because the same pitfall likely affects other long-range spin-wave
studies. Zero-point effects are appreciable here, $\delta S=0.0555$ and
$\eps_{\mathrm{qu}}=-2.7472$, reflecting the near-degeneracy of the LT doublet.
This vertex set has, to our knowledge, never been solved for dipoles, so the
entire entry is new.

\begin{figure*}[t]
\includegraphics[width=\textwidth]{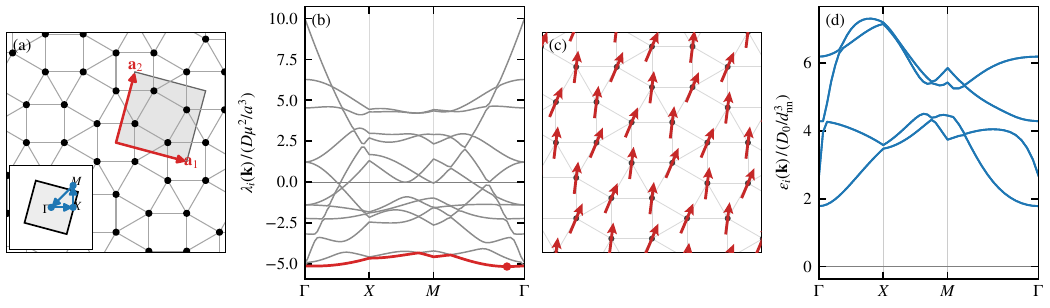}
\caption{Snub square lattice $(3^{2}.4.3.4)$, $n=4$: lattice and
Brillouin zone (a), LT bands with $\bk_0=\Gamma$ marked (b), equilibrium
configuration (c), classical spin-wave dispersion (d).
$\eps_{\min}=-2.556462$, $\delta S=0.0555$,
$\eps_{\mathrm{qu}}=-2.7472$.
Equilibrium orientations: $157.06^{\circ}$ and $172.94^{\circ}$, i.e.\ $165^{\circ}\pm7.94^{\circ}$, the incommensurate canting of Table~\ref{tab:angles}.}
\label{fig:snub_square}
\end{figure*}
\subsection{Snub hexagonal $(3^{4}.6)$}

The snub hexagonal (snub trihexagonal) lattice, known in the frustrated-magnetism
literature as the \emph{maple-leaf} lattice --- the $1/7$-depleted triangular
lattice --- is chiral, with a six-site cell built on a $\sqrt{7}$ triangular
Bravais lattice. It supports the richest ground state of the entire survey. The
order is at $\bk_{0}=0$, but the strong condition fails and free minimization
yields three degenerate mirror pairs of spins carrying \emph{three} distinct
incommensurate cantings, $6.79^{\circ}$, $11.23^{\circ}$ and $13.00^{\circ}$
(Table~\ref{tab:angles}, Fig.~\ref{fig:incomm}e), with no spin lying on any
lattice edge --- in contrast to the kagome and kisrhombille, where at least
one spin is edge-pinned. All three angles are transcendental. Because the
tiling is chiral, one might expect the two enantiomorphs to differ physically;
they do not. The dipolar energy is invariant under reflection even though
magnetic moments are axial vectors, since both $\bm{m}_{\mu}\!\cdot\bm{m}_{\nu}$
and $(\bm{m}_{\mu}\!\cdot\hat{\bm r})(\bm{m}_{\nu}\!\cdot\hat{\bm r})$ acquire a
factor $(\det R)^{2}=1$ under a reflection $R$; the two enantiomorphs therefore
share the spectrum exactly, and we compute only one. The magnon spectrum (panel d) is
gapped and stable --- again only after replacing the damped summation by the
Ewald dynamical matrix --- with $\eps_{\min}=-2.482453$ and quantum corrections
$\delta S=0.0340$, $\eps_{\mathrm{qu}}=-2.6210$. The system was flagged as
unsolved in our own working notes; it is settled here, and the three-angle
canted state is, to our knowledge, entirely new.

\begin{figure*}[t]
\includegraphics[width=\textwidth]{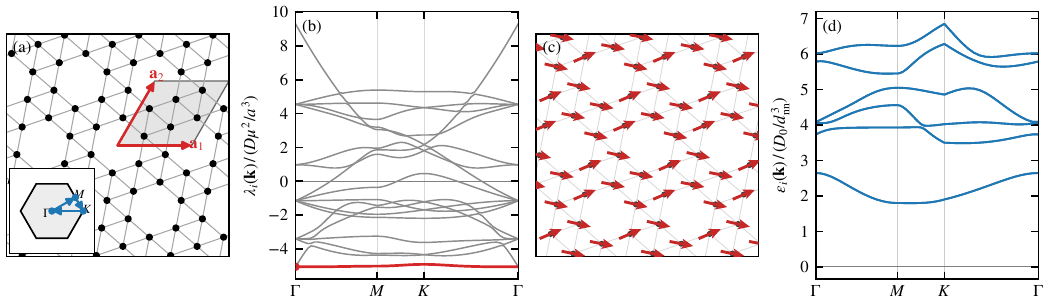}
\caption{Snub hexagonal lattice $(3^{4}.6)$, $n=6$: lattice and
Brillouin zone (a), LT bands with $\bk_0=\Gamma$ marked (b), equilibrium
configuration (c), classical spin-wave dispersion (d).
$\eps_{\min}=-2.482453$, $\delta S=0.0340$,
$\eps_{\mathrm{qu}}=-2.6210$.
Equilibrium orientations: $23.21^{\circ}$, $347.00^{\circ}$ and $348.77^{\circ}$, carrying the three independent incommensurate cantings of Table~\ref{tab:angles}.}
\label{fig:snub_hexagonal}
\end{figure*}
\section{Laves lattices}\label{sec:perlaves}

\subsection{Kisrhombille $(\mathrm{dual}(4.6.12))$}

The kisrhombille is the barycentric subdivision of the hexagonal tiling: its
vertices are the union of a triangular sublattice (hexagon centres), a
honeycomb sublattice (hexagon vertices) and a kagome sublattice (edge
midpoints), and, as shown in Appendix~\ref{app:laves}, it shares its vertex set
exactly with the deltoidal trihexagonal tiling, so the two are the same dipolar
problem. It is the Laves lattice for which an independent prior result exists,
and it provides the sharpest external validation in the survey. The order is at
$\bk_{0}=0$; four of the six spins pin exactly to lattice edges and the
remaining two form a mirror pair canted by $\pm\delta$, with a single
incommensurate angle $\delta=3.9818279307^{\circ}$
(Table~\ref{tab:angles}, Fig.~\ref{fig:incomm}b). This is precisely the
$\theta_{c}\approx4^{\circ}$ deviation reported by Batle, Bednorz and
Cerd\`a~\cite{BBC2025}, and our bulk energy agrees with their finite-cluster
value to four significant figures: $\eps_{\min}=-1.775493$ against
$-14.1977382/8=-1.774717$ once their hexagon-side length unit is converted to
$d_{\mathrm{nn}}$. The agreement is by wholly independent methods --- their
finite-cluster Levenberg--Marquardt minimization against our bulk Ewald
summation on the barycentric-subdivision construction --- and confirms both the
geometry and the value on a six-sublattice lattice. The spin-wave spectrum
(panel d) shows a nearly flat lowest band, the same corner-sharing loop
localization that flattens the kagome band (Sec.~\ref{sec:flat}), consistent
with the kisrhombille containing a kagome sublattice; the residual bandwidth
measures the long-range dipolar tail. Quantum corrections are small,
$\delta S=0.0106$ and $\eps_{\mathrm{qu}}=-1.7879$.

\begin{figure*}[t]
\includegraphics[width=\textwidth]{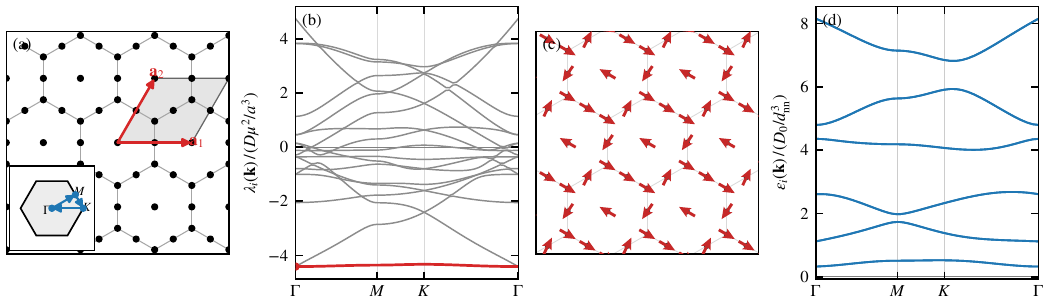}
\caption{Kisrhombille lattice $(\mathrm{dual}(4.6.12))$, $n=6$: lattice and
Brillouin zone (a), LT bands with $\bk_0=\Gamma$ marked (b), equilibrium
configuration (c), classical spin-wave dispersion (d).
$\eps_{\min}=-1.775493$, $\delta S=0.0106$,
$\eps_{\mathrm{qu}}=-1.7879$.
Equilibrium orientations: $150^{\circ}$ and $330^{\circ}$ (edge-aligned) together with $56.02^{\circ}$ and $243.98^{\circ}$, the latter canted by $3.98^{\circ}$ from the $60^{\circ}$ edge.}
\label{fig:kisrhombille}
\end{figure*}
\subsection{Floret pentagonal $(\mathrm{dual}(3^{4}.6))$}

The floret pentagonal lattice, the dual of the snub hexagonal tiling, has a
nine-site cell on a $\sqrt{21}$ triangular Bravais lattice, with all basis
sites at exact multiples of $1/21$ --- an arithmetic regularity we exploit for
high-precision geometry. Its six-petalled ``florets'' occasionally earn it a
zoological nickname in the tiling literature (variants of \emph{ladybug} are
used for pentagonal duals of this family). The order is at the zone edge $M$,
giving a $36$-site magnetic cell. The ground state is canted;
$\eps_{\min}=-1.680367$. As with the star lattice, the large magnetic cell
makes the quantum computation demanding --- a $72\times72$ Bogoliubov problem
at each wavevector --- but Brillouin-zone integration on grids up to
$12\times12$ gives fully converged values $\delta S=0.0098$ and
$\eps_{\mathrm{qu}}=-1.699812$, stable to five digits from the $5\times5$ grid
onward. Chirality of the parent
snub tiling is invisible to the dipolar energy, for the reflection-invariance
reason spelt out for the snub hexagonal lattice, so only one enantiomorph need
be computed. All branches (panel d) are gapped. No prior dipolar
treatment of this vertex set is known to us; the ground state, spectrum and
quantum estimates are new. Panel (c) shows a single magnetic cell, which
already displays the characteristic pinwheel arrangement of the floret motif.

\begin{figure*}[t]
\includegraphics[width=\textwidth]{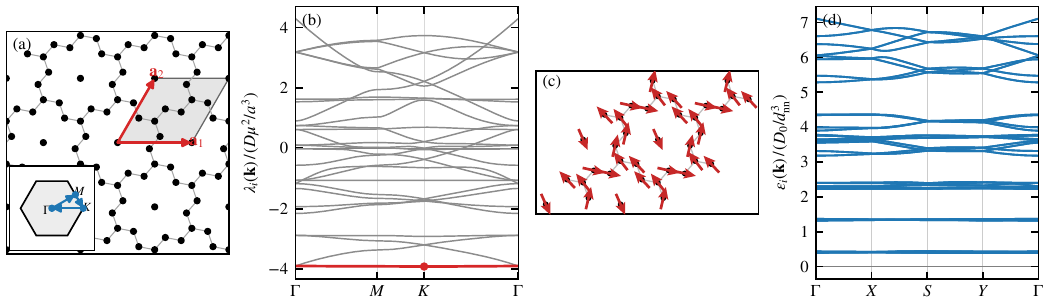}
\caption{Floret pentagonal lattice $(\mathrm{dual}(3^{4}.6))$, $n=9$: lattice and
Brillouin zone (a), LT bands with $\bk_0=M$ marked (b), equilibrium
configuration (c), classical spin-wave dispersion (d).
$\eps_{\min}=-1.680367$, $\delta S=0.0098$,
$\eps_{\mathrm{qu}}=-1.699812$.
Equilibrium orientations: $76.79^{\circ}$, $113.71^{\circ}$, $127.89^{\circ}$, $132.91^{\circ}$, $167.63^{\circ}$ --- five distinct incommensurate orientations on the $2\times2$ magnetic cell.}
\label{fig:floret_pentagonal}
\end{figure*}
\subsection{Cairo pentagonal $(\mathrm{dual}(3^{2}.4.3.4))$}

The Cairo pentagonal lattice --- the street-paving tiling of irregular
pentagons --- is the most experimentally topical member of the family and
yields the survey's only ground state with \emph{two} independent
incommensurate angles. The order is at $\bk_{0}=0$ on the six-site cell; the
strong condition fails and free minimization gives a canted state with two
distinct cantings, $\delta_{1}=3.5724^{\circ}$ and $\delta_{2}=5.5897^{\circ}$
(Table~\ref{tab:angles}, Fig.~\ref{fig:incomm}d), both transcendental. The
magnon spectrum (panel d) is gapped and stable throughout the zone, with
$\eps_{\min}=-1.324840$ and quantum corrections $\delta S=0.0058$,
$\eps_{\mathrm{qu}}=-1.3365$. What makes this lattice significant is the
contrast with its extensively studied Ising counterpart. The dipolar Cairo
lattice has been realized experimentally as an artificial spin system of
Ising-type nanomagnets by Saccone and co-workers~\cite{Saccone2019}, who found
strong geometrical frustration, a dominance of short-range correlations and an
\emph{absence} of long-range order; the finite-size Ising problem has also been
analysed exhaustively by Makarova and co-workers~\cite{Makarova2021}, who
emphasised the sensitivity of the ground state to the lattice parameter. For
freely rotating (XY or Heisenberg) dipoles we find, by contrast, a
well-defined, stable, long-range-ordered ground state. This qualitative
difference --- disordered for Ising moments, ordered and doubly canted for
continuous moments --- singles out the Cairo lattice as the most discriminating
experimental target of the family: replacing the elongated Ising islands of
Ref.~\cite{Saccone2019} by the circular discs used in the square-lattice
realization of Ref.~\cite{Leo2018} should reveal the ordered canted state
predicted here, providing a direct test of the role of spin dimensionality in
frustrated dipolar magnetism. Every result reported for this vertex set is new.

\begin{figure*}[t]
\includegraphics[width=\textwidth]{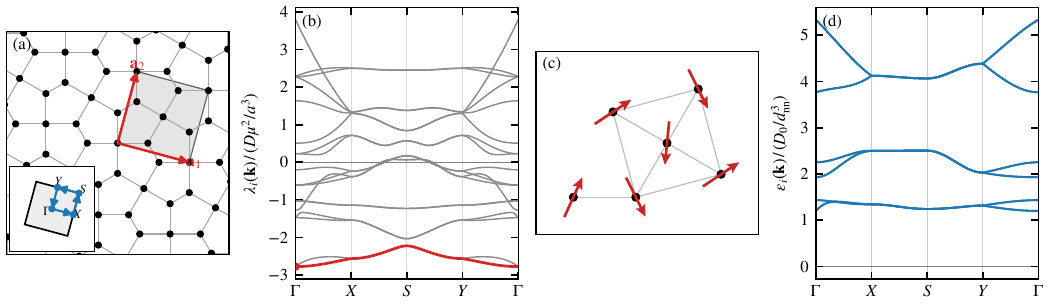}
\caption{Cairo pentagonal lattice $(\mathrm{dual}(3^{2}.4.3.4))$, $n=6$: lattice and
Brillouin zone (a), LT bands with $\bk_0=\Gamma$ marked (b), equilibrium
configuration (c), classical spin-wave dispersion (d).
$\eps_{\min}=-1.324840$, $\delta S=0.0058$,
$\eps_{\mathrm{qu}}=-1.3365$.
Equilibrium orientations: $174.41^{\circ}$, $206.43^{\circ}$, $303.57^{\circ}$, $335.59^{\circ}$, carrying the two independent cantings $3.57^{\circ}$ and $5.59^{\circ}$ of Table~\ref{tab:angles}.}
\label{fig:cairo_pentagonal}
\end{figure*}
\subsection{Prismatic pentagonal $(\mathrm{dual}(3^{3}.4^{2}))$}

The prismatic pentagonal lattice, dual of the elongated triangular tiling,
has a compact three-site cell and yields the simplest of the Laves ground
states: a collinear three-sublattice arrangement at $\bk_{0}=0$, with
$\eps_{\min}=-1.462664$. It shares with the elongated triangular lattice the
distinction of an oblique primitive cell, which we treat with the same
Brillouin-zone construction. The magnon spectrum (panel d) is gapped, with
three pairs of branches, and the zero-point corrections stay modest,
$\delta S=0.0097$ and $\eps_{\mathrm{qu}}=-1.4804$. It was treated by
one of us in Ref.~\cite{BatleAnnPhys118}; the present bulk state is consistent
with the configuration reported there, and this work supplies the bulk energy,
the full magnon spectrum and the quantum corrections. Although it is the least
structurally elaborate of the pentagonal Laves lattices, it completes the set
and, by its collinear order, provides a baseline against which the canted cairo
and floret states are measured.

\begin{figure*}[t]
\includegraphics[width=\textwidth]{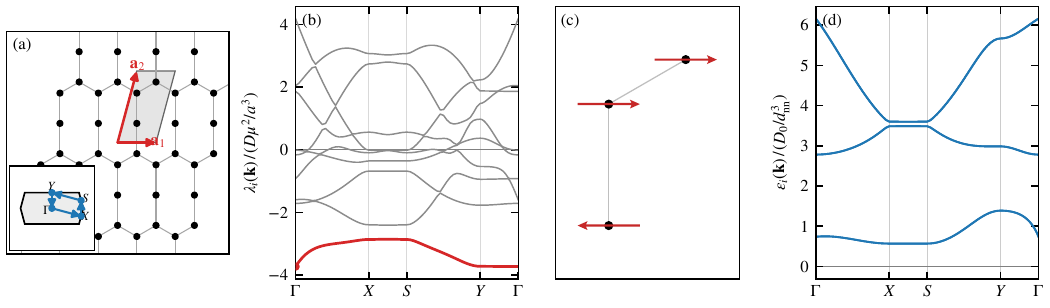}
\caption{Prismatic pentagonal lattice $(\mathrm{dual}(3^{3}.4^{2}))$, $n=3$: lattice and
Brillouin zone (a), LT bands with $\bk_0=\Gamma$ marked (b), equilibrium
configuration (c), classical spin-wave dispersion (d).
$\eps_{\min}=-1.462664$, $\delta S=0.0097$,
$\eps_{\mathrm{qu}}=-1.4804$.
Equilibrium orientations: $0^{\circ}$ and $180^{\circ}$, a collinear stripe.}
\label{fig:prismatic_pentagonal}
\end{figure*}

\section{Emergent phenomenology}\label{sec:discussion}

The fifteen preceding case studies were solved within a single formalism:
Ewald summation of the dipolar tensor, Luttinger--Tisza (LT) analysis,
unconstrained ground-state reconstruction, and linear spin-wave theory with
Holstein--Primakoff corrections. Treated in isolation, each entry answers a
narrow question about one vertex set. Treated as a controlled sample---in
which the interaction is fixed and only the lattice geometry varies---the
collection exposes regularities that no single lattice reveals. We extract
them here. Throughout, we distinguish what the sample establishes as an exact
dichotomy from what it supports only as a numerical trend, and we report the
correlations that fail as carefully as those that hold.

\begin{figure*}[t]
\includegraphics[width=\textwidth]{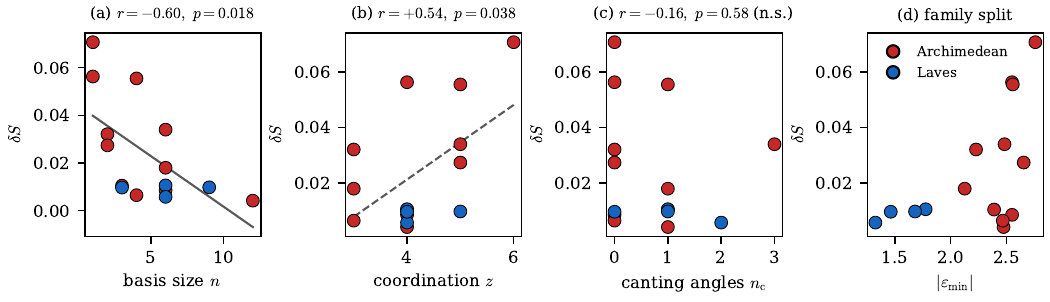}
\caption{Cross-lattice analysis of the quantum correction $\delta S$.
(a)~$\delta S$ versus basis size $n$: the one robust structural driver
($r=-0.60$, $p=0.018$), with a larger cell diluting the soft-mode zero-point
weight across more magnon branches. (b)~$\delta S$ versus coordination number
$z$: a weaker correlate whose confidence interval crosses zero (dashed fit).
(c)~$\delta S$ versus the number of independent canting angles: no relation
($r=-0.16$, $p=0.58$)---frustration does \emph{not} drive the quantum
correction. (d)~The Archimedean/Laves family split in the
$(|\varepsilon_{\min}|,\delta S)$ plane, showing that the
$\delta S$--$|\varepsilon_{\min}|$ correlation is partly a family confounder.
Red: Archimedean; blue: Laves.}
\label{fig:xlquantum}
\end{figure*}

\begin{figure*}[t]
\includegraphics[width=\textwidth]{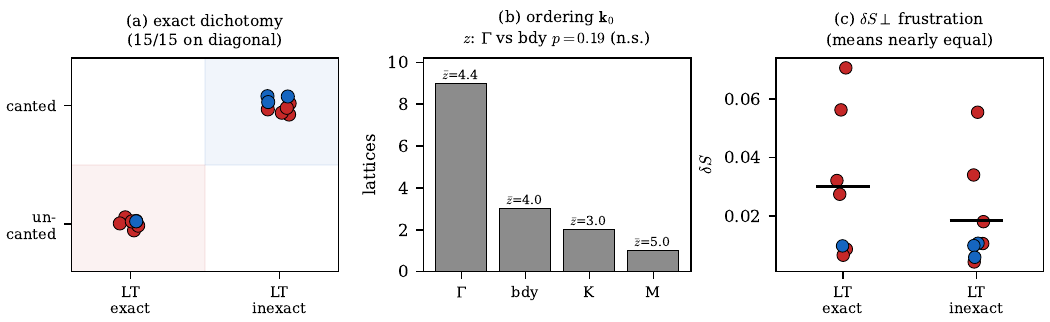}
\caption{The frustration axis. (a)~Luttinger--Tisza class versus canting: all
fifteen lattices fall on the diagonal (LT-exact\,$\leftrightarrow$\,uncanted,
LT-inexact\,$\leftrightarrow$\,canted), an exact dichotomy with no exceptions.
(b)~Ordering-vector distribution with the mean coordination $\bar z$ at each
$\mathbf{k}_0$; the $\Gamma$-versus-boundary difference in $z$ is not
significant ($p=0.19$). (c)~$\delta S$ stratified by LT class: the two class
means (bars) nearly coincide, so the fluctuation magnitude is independent of
the frustration classification.}
\label{fig:xlclassify}
\end{figure*}

\begin{table*}[t]
\caption{Cross-lattice correlation tests over the fifteen lattices ($n_{\rm c}>0$ subset where noted). Pearson $r$, two-sided $p$, and a $95\%$ bootstrap confidence interval ($5000$ resamples). Significance at $p<0.05$ is marked $\star$.}
\label{tab:xlstats}
\begin{ruledtabular}
\begin{tabular}{lcccl}
relation & $r$ & $p$ & $95\%$ CI & status \\
\hline
$\delta S$ vs basis size $n$ & $-0.60$$\star$ & $0.018$ & $[-0.80,-0.29]$ & driver of quantum corrections \\
$\delta S$ vs $|\varepsilon_{\min}|$ & $+0.55$$\star$ & $0.034$ & $[+0.26,+0.80]$ & partly family confounded \\
$\delta S$ vs coordination $z$ & $+0.54$$\star$ & $0.038$ & $[-0.07,+0.85]$ & weak; CI crosses 0 \\
$\delta S$ vs canting count $n_{\rm c}$ & $-0.16$ & $0.578$ & $[-0.63,+0.36]$ & no relation (dead) \\
$\delta S$ vs $\varepsilon_{\rm qu}$ lowering & $-0.86$$\star$ & $0.000$ & $[-0.99,-0.58]$ & consistency check \\
canting $n_{\rm c}$ vs basis (canted only) & $-0.10$ & $0.810$ & $[-0.55,+0.55]$ & dead within canted set \\
\end{tabular}
\end{ruledtabular}
\end{table*}

\begin{figure*}[t]
\includegraphics[width=\textwidth]{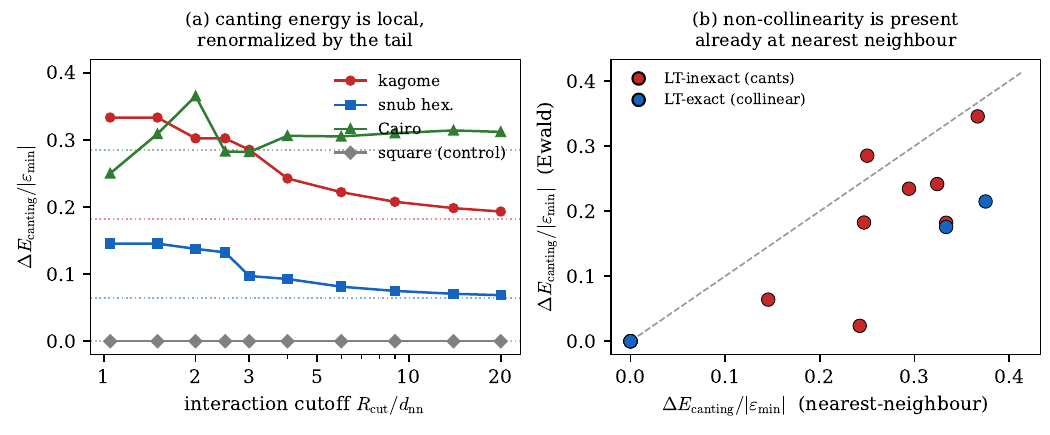}
\caption{The two-part origin of canting, from a hard interaction cutoff
$R_{\rm cut}$. (a)~Energy gained by allowing non-collinear order,
$\Delta E_{\rm canting}/|\varepsilon_{\min}|$, versus $R_{\rm cut}$ for
representative canted lattices and a collinear control (square). The gain is
already nonzero at nearest-neighbour range and is renormalized---here
reduced---toward its converged (Ewald) value, marked by the dotted lines, as
the cutoff grows: the non-collinearity is a \emph{local} property, not one
switched on by the tail. (b)~Energy gain at nearest-neighbour range versus at
full range, for all fifteen lattices; points on the diagonal keep their
non-collinearity at all ranges, while the collinear lattices sit at the origin.
The frustration is present locally; the long-range tail fixes the precise
(transcendental) angle. Red: LT-inexact (canting); blue: LT-exact.}
\label{fig:xldriving}
\end{figure*}

\subsection{Geometry controls the exactness of the Luttinger--Tisza
construction}\label{sec:vi-lt}

The LT method minimizes the dipolar tensor in reciprocal space without
enforcing the unit-length spin constraint on each sublattice. When the
resulting soft mode already satisfies that constraint site by site (the
\emph{strong} condition), the ground state is obtained directly; when it does
not, the physical ground state must be built by constrained nonlinear
minimization. Across the family the strong condition partitions the fifteen
lattices into two disjoint classes with no intermediate cases:

\begin{itemize}
\item \textbf{Class~I (LT-exact).} The strong condition holds and the ground
state follows immediately from the soft mode. This class comprises the
triangular, square, honeycomb, rhombitrihexagonal, truncated-square,
elongated-triangular and prismatic-pentagonal lattices.

\item \textbf{Class~II (LT-inexact).} The strong condition fails; the ground
state requires constrained minimization. This class comprises the kagome,
truncated-hexagonal, truncated-trihexagonal, snub-square, snub-hexagonal,
kisrhombille, floret-pentagonal and Cairo-pentagonal lattices.
\end{itemize}

The partition is not a matter of basis size. The twelve-site
truncated-trihexagonal cell is LT-inexact, whereas the six-site
rhombitrihexagonal and prismatic-pentagonal cells are LT-exact; the
four-site snub square is inexact while the six-site rhombitrihexagonal is
exact. What the two classes track instead is whether the lattice contains
sublattices with inequivalent local dipolar environments that a single
condensation wavevector cannot simultaneously optimize. Bravais nets and
decorations whose sites are all related by the point group fall in Class~I;
lattices built from corner-sharing polygons of mixed parity, or from
inequivalent vertex orbits, fall in Class~II. We state this as the operative
geometric predictor in our sample rather than as a theorem: with fifteen
lattices the classification is exact as reported, but a decisive structural
statement would require either a proof or a larger family
\cite{LuttingerTisza1946,Schmidt2003,SchmidtLuban2022}.

\subsection{Canting is the real-space fingerprint of LT
inexactness}\label{sec:vi-canting}

Eight lattices possess ground states whose sublattice orientations are
transcendental angles---incommensurate with any simple fraction of $\pi$ and
fixed only by the full lattice sum: kagome, truncated hexagonal, truncated
trihexagonal, snub square, snub hexagonal, kisrhombille, floret pentagonal and
Cairo pentagonal. For five of these the canting is carried by a small number of
independent angles, which we refine to thirty digits and collect in
Table~\ref{tab:angles}: kagome, kisrhombille and snub square each carry a
single independent canting, Cairo carries two and snub hexagonal three. The
remaining three either repeat an angle already tabulated---the truncated
hexagonal canting is the kagome angle---or, as for the floret pentagonal
lattice, distribute the canting over a larger magnetic cell; their orientations
are given in Sec.~\ref{sec:orientations}. Every one of the eight is a
Class~II (LT-inexact) lattice. In fact the correspondence in our sample is
exact and two-sided: \emph{every} LT-inexact lattice cants
($8$ of $8$) and \emph{every} LT-exact lattice does not
($7$ of $7$), with no exceptions in either direction.

This is the sharpest single statement the survey supports, and it has a
transparent interpretation. Canting is precisely the constrained solution
that appears when the unconstrained LT minimizer is inadmissible: the
inability of a single wavevector to satisfy the local constraints
simultaneously is resolved in real space by a continuous rotation of the
moments away from the lattice edges, through an angle set by the competition
between all interaction shells rather than by any local motif. The physical
content is therefore not the numerical value of any one angle but the
statement that geometric frustration, in the dipolar problem, manifests as
\emph{continuous canting determined by the long-range interaction}, and that
its presence is diagnosed exactly by the failure of the LT strong condition.

We stress that LT-exactness is not the same as collinearity. Two lattices are
LT-exact yet non-collinear with commensurate angles: the rhombitrihexagonal
lattice, whose ground state is a $60^{\circ}$ vortex (angles that are multiples
of $\pi/3$, fixed by the sixfold symmetry), and the truncated-square lattice,
whose moments lie on the two orthogonal edge axes ($0^{\circ}$ and
$90^{\circ}$, multiples of $\pi/2$). Both occupy a distinct third category:
non-collinear but commensurate, and LT-exact. The clean dichotomy is
specifically between LT-inexactness and \emph{incommensurate} canting;
commensurate non-collinear order, exemplified by these two lattices, is
compatible with LT-exactness and does not break the correspondence. Of the
fifteen lattices, five are collinear, two are commensurate non-collinear, and
eight cant at incommensurate angles.

This interpretation can be made precise, and doing so refines it. One might
guess that the canting is switched on by the long-range tail---absent at short
range, growing as the far shells are included. A direct test refutes that
simple picture and replaces it with a sharper one. We recompute the ground
state with the interaction cut at a hard radius $R_{\rm cut}$ and measure the
energy gained by allowing non-collinear order,
$\Delta E_{\rm canting}=E_{\rm collinear}-E_{\rm relaxed}\ge 0$, where
$E_{\rm collinear}$ is the best common-axis state
(Fig.~\ref{fig:xldriving}). Two facts emerge. First, for every canted lattice
$\Delta E_{\rm canting}$ is already \emph{nonzero at nearest-neighbour range}:
the non-collinearity is a local property, present in the short-range problem,
not created by the tail. For kagome, for instance, the non-collinear state lies
$33\%$ of $|\varepsilon_{\min}|$ below the best collinear state already at
$R_{\rm cut}=1.05\,d_{\rm nn}$. Second, as $R_{\rm cut}$ grows the gain is
\emph{renormalized}---for kagome, reduced---toward its converged value, and the
\emph{precise} canting angle relaxes to the transcendental value of
Table~\ref{tab:angles}. The collinear lattices, by contrast, keep
$\Delta E_{\rm canting}=0$ at every range (Fig.~\ref{fig:xldriving}b).

The driving force thus separates into two parts, and this is the resolution of
the local-versus-global question. \emph{Whether} a lattice cants is decided
locally: it is a property of the near-neighbour dipolar problem, fixed by the
coordination geometry, and coincides exactly with LT-inexactness. \emph{What}
the precise canting angle is---the transcendental value that no
nearest-neighbour model can produce---is fixed by the full long-range sum. The
frustration is local; the number is global. A nearest-neighbour model of these
lattices would already be non-collinear, but would predict the wrong angle.

The number of independent canting angles grows with the number of
symmetry-inequivalent sublattice environments rather than with the basis size
as such: it correlates only weakly with coordination number
($r \simeq +0.17$) and moderately with basis size
($r \simeq +0.48$), consistent with counting inequivalent orbits rather than
sites. A candidate finer predictor---corner-sharing of odd-membered
polygons---appears in the canted set but we do not have the statistics to
isolate it from the LT-inexactness criterion, which already accounts for the
partition completely \cite{Yu2015}.

\subsection{Ordering wavevectors}\label{sec:vi-k0}

The condensation wavevectors span $\Gamma$, $K$, $M$, and the square-lattice
boundary points $(\pi,0)$, $(0,\pi)$ and $(\pi,\pi)$. Nine of the fifteen
lattices order at $\Gamma$; the remaining six select finite-$\mathbf{k}$ order
at a Brillouin-zone boundary. The tendency in the sample is that the
higher-coordination lattices order at $\Gamma$ while the finite-$\mathbf{k}$
states occur at lower coordination (mean coordination $4.4$ at $\Gamma$ versus
$3.8$ at the boundary; Fig.~\ref{fig:xlclassify}b), but the separation is not
statistically significant ($p=0.19$) and we present it as a
weak trend rather than a rule: coordination number alone does not determine
the ordering vector, and the honeycomb ($z=3$, ordering at $K$) and
elongated-triangular ($z=5$, ordering at a boundary point) show that the
correlation is far from deterministic. The ordering vector is a property of
the full Fourier-space tensor and is not reducible to a single local
descriptor in this family \cite{DeBell2000,Maksymenko2015b}.

\subsection{Flat bands and corner-sharing topology}\label{sec:vi-flat}

Nearly dispersionless low-lying magnon branches were noted individually for
the kagome and kisrhombille lattices in Sec.~\ref{sec:flat}. Across the family
the flattest low-energy branches occur on the corner-sharing lattices,
supporting the interpretation that corner-sharing connectivity---rather than
any single lattice's accident---is the operative ingredient. The underlying
localization mechanism is not new: compactly localized loop states and the
resulting flat band are a defining feature of the kagome lattice, established
for nearest-neighbour hopping and Hubbard
models~\cite{Mielke1992,Bergman2008}, realized experimentally in kagome
metals~\cite{Lin2018,Kang2020}, and shown to persist under dipolar coupling by
Maksymenko \emph{et al.}~\cite{Maksymenko2017}. Our contribution is not the
mechanism but its systematic scope: quantifying the lowest-band width across
the corner-sharing members of the Archimedean--Laves family and confirming
that the flatness is a shared property of that structural class rather than a
kagome peculiarity. We caution that ``flat'' here is quantitative, not exact:
the long-range dipolar tail generically lifts the exact degeneracy that a
nearest-neighbour model would protect, so the branches are narrow---for the
kisrhombille the lowest LT band spans below one percent of the spectrum---rather
than strictly flat, consistent with the persistence found
in~\cite{Maksymenko2017}. The mechanism should therefore be read as a
structural tendency across the corner-sharing members, not as a protected
band-touching of the dipolar problem.

\subsection{Quantum corrections are governed by the magnon spectrum, not by
frustration}\label{sec:vi-quantum}

Every stable coplanar lattice carries a small Holstein--Primakoff moment
reduction, $\delta S$ ranging from $0.004$ (truncated trihexagonal) to
$0.071$ (triangular), and a zero-point energy lowering of a few percent
(Table~\ref{tab:master}). Because the sample is small ($15$ lattices) we test
each proposed relationship for statistical significance and report a bootstrap
confidence interval; the tests are collected in Table~\ref{tab:xlstats} and
the key relations are shown in Fig.~\ref{fig:xlquantum}. The picture that
survives this scrutiny is specific, and it overturns the natural expectation.

\emph{What drives $\delta S$.} The single robust structural predictor of the
quantum correction is the \emph{basis size}: $\delta S$ decreases with the
number of sublattices per cell ($r=-0.60$, $p=0.018$;
Fig.~\ref{fig:xlquantum}a). The relation strengthens rather than weakens when
coordination number is held fixed (partial correlation $-0.66$), and within
the LT-exact class alone it is $r=-0.81$ ($p=0.026$). The mechanism is direct:
a larger magnetic cell carries more magnon branches, so the zero-point weight
of the soft, long-wavelength acoustic mode---which dominates the
fluctuation---is spread across a denser spectrum, reducing the per-site moment
reduction. The correction is therefore a property of the excitation spectrum,
set by how the soft mode shares phase space with the rest of the band
structure \cite{Maksymenko2015b}.

\emph{Correlates that ride along.} $\delta S$ also correlates with
coordination number ($r=+0.54$, $p=0.038$; Fig.~\ref{fig:xlquantum}b) and with
the magnitude of the ground-state energy ($r=+0.55$, $p=0.034$), but we do not
read either as an independent driver. The $\delta S$--$z$ relation has a
bootstrap interval that includes zero, and the $\delta S$--$|\varepsilon_{\min}|$
relation is partly a family effect: the Archimedean lattices have both larger
$|\varepsilon_{\min}|$ (mean $2.48$ versus $1.56$) and larger $\delta S$ (mean
$0.029$ versus $0.009$) than the Laves lattices, so a correlation between the
two partly encodes the Archimedean/Laves split rather than a physical law
(Fig.~\ref{fig:xlquantum}d). Basis size, which does \emph{not} separate by
family (mean coordination $4.18$ versus $4.25$), is the cleaner variable and
the one we assert.

\emph{What does \emph{not} drive $\delta S$.} The natural expectation---that
the frustrated, canted lattices should fluctuate most---is false in this
family, and the failure is worth stating explicitly because it is itself a
result. The moment reduction is uncorrelated with the number of canting angles
($r=-0.16$, $p=0.58$; Fig.~\ref{fig:xlquantum}c), and the mean $\delta S$ of
the LT-exact class ($0.030$) slightly \emph{exceeds} that of the frustrated
LT-inexact class ($0.019$), the opposite of the expected ordering
(Fig.~\ref{fig:xlclassify}c). Geometric frustration, as diagnosed by canting,
does not enhance the quantum correction here. The reason is that canting in
these lattices is a classical, energetic rearrangement fixed by the long-range
sum; it reshapes the ground state but does not by itself soften the magnon
spectrum, which is what $\delta S$ actually measures. A single strong
correlation remains as a pure consistency check: $\delta S$ and the fractional
zero-point energy lowering track each other tightly ($r=-0.86$,
$p<10^{-3}$), as they must, since both measure the same underlying fluctuation
amplitude.

\emph{Two orthogonal axes.} Combining Secs.~\ref{sec:vi-lt}--\ref{sec:vi-quantum},
the family organizes along two statistically independent axes. The first is a
frustration axis: the LT strong condition splits the lattices into exact and
inexact classes, and this split coincides exactly with the absence or presence
of transcendental canting. The second is a fluctuation axis: the magnitude of
the quantum correction is set by the magnon spectrum through the basis size. A
lattice's position on the first axis does not predict its position on the
second (Fig.~\ref{fig:xlclassify}c). The paper's classification is thus
genuinely two-dimensional, and coordination number---often invoked as a proxy
for magnetic complexity---cleanly determines neither: it does not fix the
ordering vector (Sec.~\ref{sec:vi-k0}) nor the canting count
(Sec.~\ref{sec:vi-canting}), and it enters the quantum correction only as a
confounded correlate of basis size.

\subsection{Synthesis and experimental implications}\label{sec:vi-synthesis}

Two organizing principles survive the sample, and they are independent. The
first is exact within the family: the Luttinger--Tisza strong condition
partitions the fifteen lattices into LT-exact and LT-inexact classes, and this
partition coincides exactly with the absence or presence of transcendental
canting. This first principle has an identifiable driving force, which the
cutoff experiment of Fig.~\ref{fig:xldriving} separates into two parts:
\emph{whether} a lattice cants is a local property of the near-neighbour
dipolar problem, fixed by the coordination geometry and coinciding exactly with
LT-inexactness, while the \emph{precise} transcendental angle is fixed by the
full long-range sum. The frustration is local; the number is global. In both
respects it is diagnosed in reciprocal space by LT inexactness and realized in
real space by continuous canting. The second principle is a robust
trend: the magnitude of the quantum corrections is set by the magnon spectrum
(through the basis size), not by frustration, so that the two classifications
are orthogonal. Coordination number, often invoked as a proxy for magnetic
complexity, predicts neither cleanly---it correlates only weakly with the
ordering vector and does not determine the canting count.

For experiment, the harmonic analysis presented here is complete in a specific
sense: it provides the full linear spin-wave excitation spectrum for every
Archimedean and Laves dipolar magnet, each a falsifiable prediction for
inelastic neutron scattering or for artificial dipolar arrays
\cite{Skjaervo2020,Pac2025}. The honeycomb
case already has a direct realization in the van der Waals magnet
ErBr$_3$ \cite{Wessler2022}, whose sub-Kelvin non-collinear order and Dirac
magnon cones match the continuously degenerate honeycomb ground state found
here; the square-lattice stripe order has been observed in artificial dipolar
nanodisc arrays \cite{Leo2018}. The Class~II lattices---kagome, the snubs,
Cairo and kisrhombille---are, to our knowledge, the states where the present
predictions are most distinctive, since their transcendental canting angles
are fixed by the long-range interaction and would not arise in any
nearest-neighbor description; a disc-based realization of the Cairo array is
proposed as the sharpest discriminating experiment
\cite{Saccone2019,Makarova2021}.

\section{Conclusions}\label{sec:conclusions}

We have solved the classical dipolar problem on all fifteen distinct vertex
sets of the Archimedean and Laves tilings within a single formalism: Ewald
summation of the dipolar tensor, Luttinger--Tisza analysis to locate the
ordering wavevector, unconstrained reconstruction of the ground state on the
commensurate magnetic cell, linear spin-wave theory, and Holstein--Primakoff
quantum corrections. For each lattice we report the ground-state energy, the
ordering vector, the full set of equilibrium orientations, the magnon
spectrum, the moment reduction $\delta S$ and the zero-point-corrected energy.
Several of these vertex sets had no dipolar solution in the literature, and
for those the ground state, its spectrum and its quantum corrections are new;
where prior results exist---the kagome canting of Ref.~\cite{MCM2015}, the
kisrhombille deviation of Ref.~\cite{BBC2025}, the finite-cluster energies of
Ref.~\cite{BatleAnnPhys118}---the present bulk values agree with them and
extend them to the harmonic and quantum levels.

The fifteen solutions fall into three classes, and the classification is the
principal structural result of the survey. Five lattices order collinearly,
two order non-collinearly at angles commensurate with the lattice symmetry,
and eight cant at transcendental angles fixed by the full lattice sum and
possessing no closed form; we report the latter to thirty digits. The
commensurate class deserves emphasis because it is easy to overlook: the
rhombitrihexagonal ground state is a $60^{\circ}$ vortex and the
truncated-square ground state places its moments on the two orthogonal edge
axes, so that non-collinearity and incommensurability are genuinely distinct
properties rather than two names for the same thing. Within our sample the
correspondence between the failure of the Luttinger--Tisza strong condition and
the appearance of incommensurate canting is exact and two-sided, with no
exceptions in either direction: every lattice whose unconstrained soft mode
violates the unit-length constraint cants at a transcendental angle, and every
lattice whose soft mode satisfies it does not. Commensurate non-collinear
order, by contrast, is entirely compatible with an exact Luttinger--Tisza
construction. We state this as an exact property of the sample rather than as a
theorem, since fifteen lattices cannot settle a general structural claim, but
it is sharp enough to serve as a practical diagnostic: a reciprocal-space test
predicts a real-space signature.

The origin of that canting separates cleanly into a local and a global part,
and disentangling them corrects a natural but mistaken expectation. By cutting
the dipolar interaction at a finite radius and following the energy gained by
allowing non-collinear order, we find that for every canted lattice this gain
is already nonzero at nearest-neighbour range: whether a lattice cants at all
is decided by the local coordination geometry, not by the tail. What the
long-range interaction does is fix the precise angle, renormalising it as the
cutoff grows until it reaches the transcendental value we tabulate. The
frustration is therefore local and the number is global, and a
nearest-neighbour model of any of these lattices would produce a non-collinear
state with the wrong angle rather than no canting at all.

A second, independent organising principle governs the quantum corrections,
and it is not the one a reader would guess. The moment reduction $\delta S$ is
uncorrelated with the number of independent canting angles and is, if anything,
slightly larger in the unfrustrated class than in the frustrated one, so
geometric frustration does not enhance quantum fluctuations in this family.
What $\delta S$ tracks instead is the structure of the magnon spectrum, and in
particular the basis size: larger magnetic cells carry more branches, the
zero-point weight of the soft acoustic mode is spread over a denser spectrum,
and the per-site correction falls. The two axes along which the family
organises---frustration on the one hand, fluctuation magnitude on the
other---are thus statistically independent, and the natural single-parameter
summary, coordination number, predicts neither cleanly: it does not fix the
ordering vector, it does not determine the canting count, and it enters the
quantum correction only as a confounded correlate of basis size. Given the
sample size we present the correlations with their significance levels and
confidence intervals rather than as laws, and we report the hypotheses that
fail as carefully as those that survive.

The corner-sharing lattices host the narrowest low-energy magnon branches, the
kisrhombille lowest band spanning under one percent of the total spectral width
and the kagome band some six percent, against twenty to thirty percent for the
non-corner-sharing members. The localisation mechanism is the familiar one of
compactly localised loop states, long established for hopping and Hubbard
models on the kagome lattice and observed spectroscopically in kagome metals,
and shown to persist under dipolar coupling; what we add is not the mechanism
but its systematic scope, namely that narrow low-energy branches are a shared
property of the corner-sharing structural class across the whole
Archimedean--Laves family, in the classical dipolar magnon spectrum rather than
in an electronic band. The residual width is a quantitative measure of the
long-range tail and the bands are narrow rather than strictly flat.

Several directions follow naturally. The harmonic analysis presented here is
complete for this family, which makes each spectrum a falsifiable prediction
for inelastic neutron scattering or for artificial dipolar arrays; the
incommensurately canted lattices are the sharpest discriminators, since their
angles cannot arise in any nearest-neighbour description. An applied field
would probe the degenerate manifolds directly, and the triangular and honeycomb
cases, whose ground states are continuous families rather than isolated states,
are the natural place to look for order-by-disorder selection. Finite
temperature is untouched here and is where the flat and near-flat branches
should matter most, since a narrow band of low-lying modes carries a large
entropy at small energy cost. Competition between the dipolar tail and a
nearest-neighbour exchange is a further axis: our cutoff analysis indicates
that the non-collinearity would survive such a competition while the precise
angle would shift, which is an experimentally accessible signature. Finally,
the floret pentagonal lattice is the member whose ground state is hardest to
locate numerically, its thirty-six-spin magnetic cell requiring many
independent restarts before the minimum is found, and it is the case we would
single out for a dedicated study.

\begin{acknowledgments}
J.B. appreciates fruitful discussions with Maria Vallespir-Socias,
Joana Rossell\'o, Margalida Batle, Maria del Mar Batle, and Regina Batle.
J.B. received no funding for the present work. This work was desk-rejected by PRB.
\end{acknowledgments}

\vskip 0.5cm

\noindent {\bf Data Availability}
\vskip 0.5cm
\noindent Data will be made available on reasonable request.


\appendix

\section{The dipolar tensor and the energy functional}
\label{app:tensor}

Each site carries a dipole $\bm{m}_{i}=\mu\,\hat{\bm{m}}_{i}$ with
$\|\hat{\bm{m}}_{i}\|=1$. The pair energy is
\begin{equation}
E_{u,v}=D\left(\frac{\bm{m}_{u}\cdot\bm{m}_{v}}{r_{uv}^{3}}
-\frac{3(\bm{m}_{u}\cdot\bm{r}_{uv})(\bm{m}_{v}\cdot\bm{r}_{uv})}{r_{uv}^{5}}\right),
\label{eq:app_pair}
\end{equation}
with $D=\mu_{0}/4\pi$ (magnetic) or $1/4\pi\epsilon_{0}$ (electric). Factoring
out $\mu^{2}$ and writing $\hat{\bm{r}}=\bm{r}_{uv}/r_{uv}$,
Eq.~\eqref{eq:app_pair} becomes the quadratic form
$E_{u,v}=D\mu^{2}\,\hat{\bm{m}}_{u}^{\mathsf{T}}\bm{J}_{uv}\hat{\bm{m}}_{v}$ with
\begin{equation}
\bm{J}_{ij}=\frac{1}{\|\bm{r}_{ij}\|^{3}}
\left(\bm{I}-3\,\frac{\bm{r}_{ij}\otimes\bm{r}_{ij}}{\|\bm{r}_{ij}\|^{2}}\right),
\label{eq:app_J}
\end{equation}
where $\otimes$ is the outer product.\footnote{Equation~(11) of
Ref.~\cite{BBC2025} prints this as $\bm{r}_{ij}\times\bm{r}_{ij}$; the dyad
$\otimes$ is meant, a cross product being identically zero and not a matrix.}
All energies below are in units of $D\mu^{2}/a^{3}$, i.e.\ we set
$D=\mu=a=1$ and restore units only where stated.

The eigenvalues of $\bm{J}_{ij}$ are $(-2,+1,+1)/r_{ij}^{3}$, with the
$-2$ belonging to $\hat{\bm{r}}_{ij}$: two dipoles are most bound when
aligned head-to-tail along their separation. This is the elementary fact
behind the bound $E_{\min}\ge-2\sum_{i<j}1/r_{ij}^{3}$ of
Ref.~\cite{BBC2025}, Eq.~(2).

\section{The Luttinger--Tisza matrix and its Fourier resolution}
\label{app:LT}

\subsection{Real-space form}

Let $T(i)$ denote the images of site $i$ under the translation group of the
lattice. With $\hat{\bm{m}}_{i}=\hat{\bm{m}}_{i'}$ for all $i'\in T(i)$, the
sum in Eq.~\eqref{eq:app_pair} collapses onto one cell and the energy per
dipole is \cite{BBC2025}
\begin{equation}
E=\frac{1}{2n}\sum_{i,j=1}^{n}\hat{\bm{m}}_{i}^{\mathsf{T}}\bm{A}_{ij}\hat{\bm{m}}_{j},
\qquad
\bm{A}_{ij}=\!\!\sum_{j'\in T(j),\,j'\neq i}\!\!\bm{J}_{ij'},
\label{eq:app_A}
\end{equation}
with $n$ the number of dipoles per cell. Assembling the $\bm{A}_{ij}$ into
the $nd\times nd$ matrix $\hat{\bm{A}}$ ($d=2$ for XY, $d=3$ for Heisenberg
dipoles) gives $E=(1/2n)\,\hat{\bm{m}}^{\mathsf{T}}\hat{\bm{A}}\,\hat{\bm{m}}$.
Diagonalizing, $\hat{\bm{A}}\hat{\bm{x}}_{k}=\lambda_{k}\hat{\bm{x}}_{k}$ with
$\|\hat{\bm{x}}_{k}\|=\sqrt{n}$, and expanding
$\hat{\bm{m}}=\sum_{k}a_{k}\hat{\bm{x}}_{k}$,
\begin{equation}
E=\frac{1}{2}\sum_{k=1}^{nd}\lambda_{k}a_{k}^{2}.
\label{eq:app_Ek}
\end{equation}
The \emph{weak} condition $\sum_{k}a_{k}^{2}=\mu^{2}$ fixes the overall
normalization; the \emph{strong} condition
$\|\sum_{k}a_{k}\bm{x}_{k}^{i}\|=\mu$ for every $i=1,\dots,n$ enforces the
hard unit-length constraint site by site. Minimizing
Eq.~\eqref{eq:app_Ek} subject to the weak condition alone puts all the
weight on the lowest mode,
\begin{equation}
E_{\min}=\tfrac{1}{2}\lambda_{\min}\mu^{2}
\label{eq:app_Emin}
\end{equation}
which is a rigorous lower bound in general and the exact ground-state
energy whenever the strong condition is met.

\subsection{Fourier resolution}

Equation~\eqref{eq:app_A} fixes a cell in advance. Writing instead the sites
as $\bm{r}_{\bm{R},\mu}=\bm{R}+\bm{\tau}_{\mu}$ with $\bm{R}$ in the Bravais
lattice $\Lambda$ and $\bm{\tau}_{\mu}$ the $n$ basis positions, define
\begin{equation}
A^{ab}_{\mu\nu}(\bm{k})=\sideset{}{'}\sum_{\bm{R}\in\Lambda}
J^{ab}(\bm{R}+\bm{\tau}_{\mu}-\bm{\tau}_{\nu})\,e^{i\bm{k}\cdot\bm{R}},
\label{eq:app_Ak}
\end{equation}
the prime excluding $\bm{R}=0$ when $\mu=\nu$. Then $\hat{\bm{A}}=\bm{A}(\bm{k}=0)$
for the chosen cell, and
\begin{equation}
\lambda_{\min}=\min_{\bm{k}\in\mathrm{BZ}}\lambda_{\min}\!\left[\bm{A}(\bm{k})\right],
\end{equation}
attained at the ordering vector $\bm{k}_{0}$. An ordering vector
$\bm{k}_{0}\neq0$ of the primitive cell reappears as a $\bm{k}=0$ mode of any
supercell commensurate with $\bm{k}_{0}$; Eq.~\eqref{eq:app_Emin} is
unchanged. We work with $\bm{A}(\bm{k})$ because it locates $\bm{k}_{0}$
without guessing the magnetic cell, and because the spin-wave problem of
Appendix~\ref{app:sw} requires $\bm{k}$ in any case.

\subsection{Reconstruction of the real configuration}
\label{app:recon}

Let $\bm{v}$ be the lowest eigenvector of $\bm{A}(\bm{k}_{0})$. Consider a
supercell commensurate with $\bm{k}_{0}$ and the trial vector
$\psi_{\bm{R},\mu}=u_{\mu}e^{i\bm{k}_{0}\cdot\bm{R}}$. Using
Eq.~\eqref{eq:app_Ak} and the fact that, as $\bm{R}'$ runs over the cell
representatives and $\bm{R}_{\mathrm{sc}}$ over the supercell lattice,
$\bm{R}-\bm{R}'+\bm{R}_{\mathrm{sc}}$ runs over all of $\Lambda$,
\begin{equation}
\bigl[\hat{\bm{A}}\psi\bigr]_{\bm{R},\mu}
=e^{i\bm{k}_{0}\cdot\bm{R}}\bigl[\bm{A}(-\bm{k}_{0})\,\bm{u}\bigr]_{\mu}.
\end{equation}
Hence $\psi$ is an eigenvector of $\hat{\bm{A}}$ iff $\bm{u}$ is an
eigenvector of $\bm{A}(-\bm{k}_{0})=\bm{A}(\bm{k}_{0})^{*}$, i.e.\ iff
$\bm{u}=\bm{v}^{*}$. The real ground state is therefore
\begin{equation}
\hat{\bm{m}}_{\bm{R},\mu}
=c\,\mathrm{Re}\!\left[\bm{v}_{\mu}\,e^{-i\bm{k}_{0}\cdot\bm{R}}\right]
\label{eq:app_recon}
\end{equation}
with the \emph{minus} sign in the exponent, $c$ fixed by
$\|\hat{\bm{m}}\|=1$. The opposite sign yields a configuration that is not an
eigenvector of $\hat{\bm{A}}$; the error is invisible whenever $\bm{v}$ is
real (square, triangular) and material whenever $\bm{v}$ is complex
(honeycomb).

Writing $\bm{v}_{\mu}=\bm{p}_{\mu}+i\bm{q}_{\mu}$, Eq.~\eqref{eq:app_recon}
gives $\hat{\bm{m}}\propto\bm{p}_{\mu}\cos\phi+\bm{q}_{\mu}\sin\phi$ with
$\phi=\bm{k}_{0}\cdot\bm{R}$, whose modulus is $\bm{R}$-independent iff
\begin{equation}
\|\bm{p}_{\mu}\|=\|\bm{q}_{\mu}\|,
\qquad
\bm{p}_{\mu}\cdot\bm{q}_{\mu}=0
\label{eq:app_circ}
\end{equation}
for every $\mu$ (circular polarization), or $\bm{v}_{\mu}$ is real up to a
global phase. Together with $\|\bm{v}_{\mu}\|$ independent of $\mu$,
Eq.~\eqref{eq:app_circ} is the practical form of the strong condition for a
single-mode LT state. When it fails, the ground state requires a
superposition of degenerate modes and $\lambda_{\min}/2$ is a strict lower
bound.

\section{\texorpdfstring{Two-dimensional Ewald summation of $\bm{A}(\bm{k})$}{Two-dimensional Ewald summation of A(k)}}
\label{app:ewald}

The sum \eqref{eq:app_Ak} converges only conditionally, and the accuracy of
$E_{\min}$ is set entirely by how it is evaluated. We split
$1/r=\mathrm{erfc}(\alpha r)/r+\mathrm{erf}(\alpha r)/r$ and use
$J^{ab}(\bm{x})=-\partial_{a}\partial_{b}(1/r)$.

\subsection{Real-space part}

Differentiating $\mathrm{erfc}(\alpha r)/r$ twice,
\begin{equation}
-\partial_{a}\partial_{b}\frac{\mathrm{erfc}(\alpha r)}{r}
=\delta_{ab}B(r)-x_{a}x_{b}\,C(r),
\end{equation}
\begin{align}
B(r)&=\frac{1}{r^{3}}\left[\mathrm{erfc}(\alpha r)
+\frac{2\alpha r}{\sqrt{\pi}}e^{-\alpha^{2}r^{2}}\right],\\
C(r)&=\frac{1}{r^{5}}\left[3\,\mathrm{erfc}(\alpha r)
+\frac{2\alpha r}{\sqrt{\pi}}\bigl(3+2\alpha^{2}r^{2}\bigr)e^{-\alpha^{2}r^{2}}\right].
\end{align}
For an in-plane lattice $x_{z}=0$, so the $xz$ and $yz$ blocks vanish and the
$zz$ block reduces to $B(r)$. This part converges as
$e^{-\alpha^{2}r^{2}}$.

\subsection{Reciprocal part}

With $\bm{q}=\bm{k}+\bm{G}$, $q=\|\bm{q}\|$, the planar Poisson summation of
$\mathrm{erf}(\alpha r)/r$ evaluated at $z=0$ gives the kernel
\begin{equation}
F(q)=\frac{2\pi}{q}\,\mathrm{erfc}\!\left(\frac{q}{2\alpha}\right).
\end{equation}
The in-plane second derivatives bring down $+q_{a}q_{b}$, while
\begin{equation}
\left.\partial_{z}^{2}F\right|_{z=0}
=2\pi q\,\mathrm{erfc}\!\left(\frac{q}{2\alpha}\right)
-4\sqrt{\pi}\,\alpha\,e^{-q^{2}/4\alpha^{2}},
\end{equation}
so that, with $A_{\mathrm{uc}}$ the unit-cell area,
\begin{align}
A^{ab,\mathrm{lr}}_{\mu\nu}&=\frac{1}{A_{\mathrm{uc}}}\sum_{\bm{G}}
e^{-i\bm{q}\cdot\bm{\tau}_{\mu\nu}}\,q_{a}q_{b}F(q),
\quad a,b\in\{x,y\},\\
A^{zz,\mathrm{lr}}_{\mu\nu}&=\frac{1}{A_{\mathrm{uc}}}\sum_{\bm{G}}
e^{-i\bm{q}\cdot\bm{\tau}_{\mu\nu}}
\left[4\sqrt{\pi}\alpha e^{-q^{2}/4\alpha^{2}}
-2\pi q\,\mathrm{erfc}\!\left(\frac{q}{2\alpha}\right)\right],
\end{align}
with $\bm{\tau}_{\mu\nu}=\bm{\tau}_{\mu}-\bm{\tau}_{\nu}$. The $q\to0$ term
(present only at $\bm{k}=0$, $\bm{G}=0$) is finite: the in-plane blocks
vanish as $q_{a}q_{b}/q$, and the $zz$ block tends to
$4\sqrt{\pi}\alpha/A_{\mathrm{uc}}$. This part converges as
$e^{-q^{2}/4\alpha^{2}}$.

\subsection{Self term}

For $\mu=\nu$ the reciprocal sum reinstates the $\bm{R}=0$ term excluded by
the prime in Eq.~\eqref{eq:app_Ak}. Subtracting its $r\to0$ limit gives the
correction
\begin{equation}
A^{ab,\mathrm{self}}_{\mu\mu}=-\frac{4\alpha^{3}}{3\sqrt{\pi}}\,\delta_{ab}.
\end{equation}

\subsection{Validation}

Both halves converge exponentially and the total is independent of the
splitting parameter $\alpha$, which is the internal consistency test: we
find agreement to $\lesssim10^{-14}$ over $\alpha\in[1,3]$. The external test
is an exact identity: the $zz$ block of the square lattice at $\bm{k}=0$ is an
Epstein zeta function, for which Ref.~\cite{BBC2025}, Eq.~(4), gives the
closed form
\begin{equation}
A^{zz}(\bm{0})=\sideset{}{'}\sum_{\bm{x}\in\Lambda}\frac{1}{|\bm{x}|^{3}}
=4\,\zeta\!\left(\tfrac{3}{2}\right)\beta\!\left(\tfrac{3}{2}\right)
=9.033621\ldots,
\end{equation}
with $\zeta$ the Riemann and $\beta$ the Dirichlet beta function; the Ewald
sum reproduces this to eight digits. Independently, the reconstructed
honeycomb ground state of Sec.~\ref{sec:results} evaluated by direct
pairwise summation over a $7442$-dipole patch, using neither Ewald nor the
LT construction, agrees with $\lambda_{\min}/2$ to $3\times10^{-8}$.

\section{Classical linearized spin waves}
\label{app:sw}

Around a ground state $\hat{\bm{m}}^{0}_{\mu}$ on the magnetic cell we erect
local orthonormal frames
$\{\hat{\bm{e}}^{1}_{\mu},\hat{\bm{e}}^{2}_{\mu},\hat{\bm{m}}^{0}_{\mu}\}$ and set
\begin{equation}
\hat{\bm{m}}_{\mu}=\hat{\bm{m}}^{0}_{\mu}\sqrt{1-u_{\mu}^{2}-v_{\mu}^{2}}
+u_{\mu}\hat{\bm{e}}^{1}_{\mu}+v_{\mu}\hat{\bm{e}}^{2}_{\mu}.
\end{equation}
Expanding the energy to second order in $(u,v)$, the transverse Hessian is
\begin{align}
H^{ij}_{\mu\nu}(\bm{k})
&=\hat{\bm{e}}^{i}_{\mu}\cdot\bm{A}_{\mu\nu}(\bm{k})\cdot\hat{\bm{e}}^{j}_{\nu}
-\delta_{\mu\nu}\delta_{ij}\,
\hat{\bm{m}}^{0}_{\mu}\!\cdot\bm{h}_{\mu},
\\
\bm{h}_{\mu}&=\sum_{\nu}\bm{A}_{\mu\nu}(\bm{0})\,\hat{\bm{m}}^{0}_{\nu},
\end{align}
the subtracted term being the local field that enforces
$\partial E/\partial\hat{\bm{m}}_{\mu}\parallel\hat{\bm{m}}^{0}_{\mu}$ at
equilibrium. The dynamics are the Landau--Lifshitz equations
$\dot{\hat{\bm{m}}}_{\mu}=-\hat{\bm{m}}_{\mu}\times\partial E/\partial\hat{\bm{m}}_{\mu}$;
linearized in the local frame they become
\begin{equation}
\omega\,\Psi(\bm{k})=i\,\Sigma_{y}H(\bm{k})\,\Psi(\bm{k}),
\qquad
\Sigma_{y}=\mathrm{diag}(\sigma_{y},\dots,\sigma_{y}),
\end{equation}
with $\Psi=(u_{1},v_{1},\dots,u_{n},v_{n})^{\mathsf{T}}$, or equivalently
\begin{equation}
\omega^{2}\,\Psi=\mathsf{M}(\bm{k})\,\Psi,
\qquad
\mathsf{M}=\Sigma_{y}H\Sigma_{y}H .
\end{equation}
The frequencies $\varepsilon_{i}(\bm{k})=\omega_{i}(\bm{k})$ are classical
normal modes: no Holstein--Primakoff expansion, no zero-point energy, no
quantum corrections. Real non-negative $\varepsilon_{i}$ over the whole zone
certify local stability; an $\omega^{2}<0$ branch signals that the assumed
configuration is a saddle.

For XY dipoles ($d=2$) the out-of-plane deviation is frozen, one transverse
direction is lost per site, and the count of branches is $n$ rather than
$2n$. Since the dipolar ground states of planar lattices are in-plane
(Appendix~\ref{app:LT}), the two choices share a ground state but not a
spectrum: the Heisenberg spectrum contains, in addition, the out-of-plane
branches, which are the ones gapped by the easy-plane anisotropy of the
dipolar tensor.

\section{Sector decomposition and the dimension of the dipole}
\label{app:sectors}

For a planar lattice every separation lies in the plane, $r_{z}=0$, so
Eq.~\eqref{eq:app_J} gives $J^{xz}=J^{yz}=0$ identically and $\bm{A}(\bk)$
block-diagonalizes exactly,
\begin{equation}
\bm{A}(\bk)=\bm{A}_{\parallel}(\bk)\oplus A_{zz}(\bk),
\end{equation}
into a $2n\times2n$ in-plane block and an $n\times n$ out-of-plane block, the
latter a scalar $1/r^{3}$ problem, $A_{zz}(\bk)=\sum'_{\bR}e^{i\bk\cdot\bR}/
|\bR+\btau_{\mu\nu}|^{3}$. The decoupling is exact, not approximate: we find
the coupling block to be zero to machine representation.

Consequently the classical ground state is in-plane if and only if
\begin{equation}
\min_{\bk}\lambda_{\min}\!\left[\bm{A}_{\parallel}(\bk)\right]
<\min_{\bk}A_{zz,\min}(\bk),
\end{equation}
and in that case the XY ($d=2$) and Heisenberg ($d=3$) treatments share a
ground state and the same $E_{\min}$. This inequality holds, by roughly a
factor of two, for every lattice reported here; we verify it case by case
rather than assume it, since it is what licenses comparison with the XY
results of Refs.~\cite{BC2020,BBC2025}.

The spin-wave problem is a different matter and we use $d=3$ throughout. A
dipole rigidly confined to the plane has no Landau--Lifshitz precession,
since the torque $\bm{m}\times\partial E/\partial\bm{m}$ tilts the moment out
of the plane; an XY spin-wave spectrum would require positing some other
dynamics. The out-of-plane branches are physical, and are gapped by the
easy-plane anisotropy encoded in $A_{zz}>\lambda_{\min}[\bm{A}_{\parallel}]$.

One identity is worth recording, as it looks at first like a numerical
accident: the out-of-plane minima of the triangular and honeycomb lattices
coincide to machine precision. Partition the triangular lattice into three
sublattices by $\mathbf{K}\cdot\bR \bmod 2\pi$. The class $\mathbf{K}\cdot\bR=0$ is the
triangular sublattice $\Lambda_{\sqrt3}$, with sum $a$; three-fold symmetry
makes the two remaining sums equal, $T$, so
$A_{zz}^{\triangle}(\mathbf{K})=a+(\omega+\omega^{2})T=a-T$ with
$\omega=e^{2\pi i/3}$. The honeycomb sublattices are $\Lambda_{\sqrt3}$ and a
shifted copy, so its N\'eel value at $\Gamma$ is $a-|b|=a-T$: the same
quantity.

\section{Laves vertex sets and the reduction of the survey}
\label{app:laves}

The dipolar energy depends only on the positions of the dipoles. A tiling's
bond graph -- its face-transitivity, its coordination numbers, its
distinction between three- and six-fold vertices -- does not enter
Eq.~\eqref{eq:app_pair}. Two tilings sharing a vertex set are therefore the
same dipolar problem, however different they are as tilings. This is not a
formality: it removes cases from the Laves family outright.

We build each Laves vertex set as a genuine dual, by tracing the faces of the
planar nearest-neighbour graph of the Archimedean parent and taking face
centres (for Archimedean tilings all faces are regular, so the centroid and
the circumcentre agree), and compare point sets by a scale- and
orientation-invariant signature: normalize to $d_{\mathrm{nn}}=1$ and record, for each
site whose neighbourhood lies well inside the patch, the sorted distances to
its $k$ nearest neighbours. Two collapses follow.

The \emph{rhombille} (dice) tiling, dual of the kagome, has vertices at the
kagome triangle centres and hexagon centres. The triangle centres form a
honeycomb and the hexagon centres are precisely that honeycomb's own hexagon
centres; their union is the triangular lattice. Every vertex has six
neighbours at the rhombus edge length. The degree-three and degree-six
vertices are distinguishable only as graph data, which the dipolar
Hamiltonian cannot see.

The \emph{tetrakis square} tiling, dual of $(4.8^{2})$, has vertices at the
octagon centres, a square lattice of spacing $1+\sqrt{2}$, and the square
centres, an identical lattice offset by half a diagonal. Their union is a
square lattice of spacing $(1+\sqrt{2})/\sqrt{2}$ rotated by $45^{\circ}$.

Neither tiling therefore contributes a new dipolar problem: the rhombille
reproduces the triangular lattice and the tetrakis square reproduces the
square lattice, results already reported above. The
\emph{deltoidal trihexagonal}, dual of $(3.4.6.4)$, is by contrast a genuinely
distinct vertex set with three inequivalent sites. The survey is thus not a
list of eleven Archimedean and eight Laves lattices, but a list of the
distinct vertex sets among them, which is shorter.

\bibliography{refs}

\end{document}